\documentclass[reprint,twocolumn,pre,showpacs,amsmath,amssymb,aps,superscriptaddress]{revtex4-1}

\usepackage{natbib}
\usepackage{graphicx,color,rotating}
\usepackage[latin1]{inputenc}
\usepackage{textcomp}
\usepackage{dcolumn}
\usepackage{bm}     
\usepackage{upgreek}
\usepackage{subdepth}  			
\usepackage{hyperref}
\usepackage{todo}

\usepackage{xcolor}						
\hypersetup{							
    colorlinks,
    allcolors={blue}
}

\makeatletter
\renewcommand*\env@matrix[1][*\c@MaxMatrixCols c]{%
	\hskip -\arraycolsep
	\let\@ifnextchar\new@ifnextchar
	\array{#1}}
\makeatother

\renewcommand{\div}{\nablabf \cdot}
\newcommand{\divV}{\nablaV \cdot}
\newcommand{\grad}{\nablabf \!}
\newcommand{\transp}{^\mr{T}}
\newcommand{\mr}[1]{\ensuremath{\mathrm{#1}}}
\renewcommand{\vec}[1]{\bm{#1}}
\newcommand{\ee}{\mr{e}}
\newcommand{\ii}{\mr{i}}

\newcommand{\dd}{\mr{d}}

\newcommand{\dA}{\mr{d}A}

\newcommand{\avr}[1]{\big\langle #1 \big\rangle}
\newcommand{\inv}{^{-1}}

\DeclareMathOperator{\re}{Re}

\newcommand{\iot}{{\ii\omega t}}
\newcommand{\eiot}{\ee^{-\iot}}

\newcommand{\lamSAW}{\lambda_\mr{SAW}}

\newcommand{\ve}{\varepsilon}

\newcommand{\pp}{\partial}

\newcommand{\nablabf}{\boldsymbol{\nabla}}

\newcommand{\nablaV}{\bm{\nabla}_\mr{V}}

\newcommand{\ct}{\mathcal{C}}

\newcommand{\st}{{\mathcal{S}}}

\newcommand{\etal}{\textit{et~al.}}

\newcommand{\lnb}{128$^\circ$ YX-cut lithium niobate}
\newcommand{\scap}{\!\cdot\!}

\newcommand{\AAA}{\vec{A}}

\newcommand{\BBB}{\vec{B}}

\newcommand{\CCC}{\vec{C}}

\newcommand{\ccc}{\vec{c}}

\newcommand{\cccG}{\ccc^\mr{gl}_\mr{r}}
\newcommand{\cccM}{\ccc^\mr{mt}_\mr{r}}

\newcommand{\DDD}{\vec{D}}

\newcommand{\EEE}{\vec{E}}

\newcommand{\eee}{\vec{e}}
\newcommand{\eeeG}{\eee^\mr{gl}_\mr{r}}
\newcommand{\eeeM}{\eee^\mr{mt}_\mr{r}}
\newcommand{\een}{\vec{e}}

\newcommand{\FFF}{\vec{F}}

\newcommand{\FFFrad}{\vec{F}^\mr{rad}}

\newcommand{\FFFdrag}{\FFF^{\mr{drag}}}

\newcommand{\fffrad}{\vec{f}^\mr{rad}}

\newcommand{\fSAW}{f_\mr{SAW}}

\newcommand{\JJJ}{\vec{J}}

\newcommand{\ks}{k_\mr{s}}

\newcommand{\MMM}{\vec{M}}
\newcommand{\MMMs}{\vec{M}_\sigma}

\newcommand{\nnn}{\vec{n}}

\newcommand{\PPP}{\vec{P}}
\newcommand{\PPPG}{\vec{P}^\mr{gl}}
\newcommand{\PPPM}{\vec{P}^\mr{mt}}

\newcommand{\ppI}{{p_1}}
\newcommand{\ppII}{{p_2}}

\newcommand{\RRR}{\vec{R}}

\newcommand{\rrr}{\vec{r}}

\newcommand{\SSS}{\vec{S}}

\newcommand{\uuu}{\vec{u}}

\newcommand{\vvv}{\vec{v}}

\newcommand{\vvvsl}{\vec{v}_\mr{sl}}

\newcommand{\vvvP}{\vec{v}_\mr{pt}}

\newcommand{\zerovec}{\boldsymbol{0}}

\newcommand{\oneten}{\bm{I}}

\newcommand{\Zac}{Z^\mr{ac}}
\newcommand{\Zel}{Z^\mr{el}}

\newcommand{\eps}{\epsilon}

\newcommand{\epsV}{\bm{\epsilon}_\mr{V}^\mr{}}

\newcommand{\er}{\bm{\varepsilon}_\mr{r}}
\newcommand{\erG}{\bm{\varepsilon}^\mr{gl}_\mr{r}}
\newcommand{\erM}{\bm{\varepsilon}^\mr{mt}_\mr{r}}
\newcommand{\epseps}{\epsilon}

\newcommand{\eikm}{\epsilon^\mr{}_{ik}}

\newcommand{\etafl}{\eta_\mr{fl}}

\newcommand{\kappafl}{\kappa_\mr{fl}}

\newcommand{\cfl}{c_\mr{fl}}

\newcommand{\pI}{p_1}

\newcommand{\vvvI}{\vvv_1}

\newcommand{\vvvII}{{\vvv_2}}

\newcommand{\SIC}{\textrm{C}}

\newcommand{\SIum}{\upmu\textrm{m}}

\newcommand{\SIMHz}{\textrm{MHz}}

\newcommand{\SIkg}{\textrm{kg}}
\newcommand{\SIkgm}{\textrm{kg}\:\textrm{m$^{-3}$}}
\newcommand{\SIkgpcm}{\SIkgm}
\newcommand{\SIm}{\textrm{m}}

\newcommand{\SImum}{\textrm{\textmu{}m}}

\newcommand{\SIpN}{\textrm{pN}}

\newcommand{\SIMRayl}{\textrm{MRayl}}

\newcommand{\SIpTPa}{\textrm{TPa}^{-1}}

\newcommand{\SIPas}{\textrm{Pa}\:\textrm{s}}

\newcommand{\SIs}{\textrm{s}}

\newcommand{\SIV}{\textrm{V}}

\newcommand{\beq}[1]{\begin{equation} \eqlab{#1}}
\newcommand{\eeq}{\end{equation}}
\newcommand{\bsub}{\begin{subequations}}
	\newcommand{\esub}{\end{subequations}}
\def\bal#1\eal{\begin{align}#1\end{align}}
\def\bsubal#1\esubal{\bsub \begin{align}#1\end{align} \esub}

\newcommand{\nn}{\nonumber}
\newcommand{\eqlab}[1]{\label{eq:#1}}
\renewcommand{\eqref}[1]{Eq.~(\ref{eq:#1})}
\newcommand{\eqrefnoEq}[1]{(\ref{eq:#1})}
\newcommand{\eqsref}[2]{Eqs.~(\ref{eq:#1}) and~(\ref{eq:#2})}
\newcommand{\eqsrefnoEq}[2]{(\ref{eq:#1}) and~(\ref{eq:#2})}

\newcommand{\figref}[1]{Fig.~\ref{fig:#1}}

\newcommand{\figlab}[1]{\label{fig:#1}}
\newcommand{\appref}[1]{Appendix~\ref{sec:#1}}

\newcommand{\secref}[1]{Section~\ref{sec:#1}}
\newcommand{\secsref}[2]{Sections~\ref{sec:#1} and~\ref{sec:#2}}
\newcommand{\seclab}[1]{\label{sec:#1}}
\newcommand{\tabref}[1]{Table~\ref{tab:#1}}

\newcommand{\tablab}[1]{\label{tab:#1}}
\newcommand{\citeref}[1]{Ref.~\cite{#1}}

\newcommand{\GammaF}{\Gamma_\mr{fl}}
\newcommand{\GammaS}{\Gamma_\mr{sl}}
\newcommand{\rhofl}{\rho_\mr{fl}}

\newcommand{\rhosl}{\rho_\mr{sl}}

\newcommand{\sig}{\sigmabf}

\newcommand{\sigmabf}{\boldsymbol{\sigma}}

\newcommand{\sigsl}{\sig_\mr{sl}}
\newcommand{\sigik}{\sigma_{ik}}

\newcommand{\sigV}{\sig_\mr{V}}
\newcommand{\sigVG}{\sig^{\mr{gl}}_\mr{V}}
\newcommand{\sigVM}{\sig^{\mr{mt}}_\mr{V}}

\renewcommand{\Re}{\mr{Re}}

\newcommand{\shortrange}{\delta} 			
\newcommand{\atsurface}{0}				

\newcommand{\vvvshort}{\vvv}

\newcommand{\vvvds}{\vvvshort^{\shortrange\atsurface}}
\newcommand{\vvvdsC}{\vvvshort^{\shortrange\atsurface*}}

\begin{document}

\title{Three-Dimensional Numerical Modeling of Surface Acoustic Wave Devices: \\ Acoustophoresis of Micro- and Nanoparticles including Streaming}

\author{Nils R. Skov}
\email{nilsre@fysik.dtu.dk}
\affiliation{Department of Physics, Technical University of Denmark, DTU Physics Building 309, DK-2800 Kongens Lyngby, Denmark}

\author{Prateek Sehgal}
\email{ps824@cornell.edu}
\affiliation{Sibley School of Mechanical and Aerospace Engineering, Cornell University, Ithaca, New York 14853, USA}

\author{Brian J. Kirby}
\email{kirby@cornell.edu}
\affiliation{Sibley School of Mechanical and Aerospace Engineering, Cornell University, Ithaca, New York 14853, USA}
\affiliation{Department of Medicine, Division of Hematology and Medical Oncology, Weill-Cornell Medicine, New York, New York 10021, USA}

\author{Henrik Bruus}
\email{bruus@fysik.dtu.dk}
\affiliation{Department of Physics, Technical University of Denmark, DTU Physics Building 309, DK-2800 Kongens Lyngby, Denmark}

\date{25 June 2019}

\begin{abstract}
Surface acoustic wave (SAW) devices form an important class of acoustofluidic devices, in which the acoustic waves are generated and propagate along the surface of a piezoelectric substrate. Despite their wide-spread use, only a few fully three-dimensional (3D) numerical simulations have been presented in the literature. In this paper, we present a 3D numerical simulation taking into account the electromechanical fields of the piezoelectric SAW device, the acoustic displacement field in the attached elastic material, in which the liquid-filled microchannel is embedded, the acoustic fields inside the microchannel, as well as the resulting acoustic radiation force and streaming-induced drag force acting on micro- and nanoparticles suspended in the microchannel. A specific device design is presented, for which the numerical predictions of the acoustic resonances and the acoustophoretic repsonse of suspended microparticles in 3D are successfully compared with experimental observations. The simulation provides a physical explanation of the the observed qualitative difference between devices with an acoustically soft and hard lid in terms of traveling and standing waves, respectively. The simulations also correctly predict the existence and position of the observed in-plane streaming flow rolls. The presented simulation model may be useful in the development of SAW devices optimized for various acoustofluidic tasks.
\end{abstract}

\maketitle

\section{Introduction} \seclab{Intro}

During the past decade, surface acoustic wave (SAW) devices have been developed for a multitude of different types of acoustofluidic handling of micrometer-sized particles inside closed microchannels. Examples include acoustic mixing \cite{Sritharan2006}, continuous particle or droplet focusing \cite{Shi2008, Franke2009} and separation \cite{Tan2009, Shi2009}, single-particle handling \cite{Ding2012, Tran2012}, acoustic tweezing \cite{Shi2009a, Collins2016, Riaud2017}, two-dimensional single patterning \cite{Collins2015, Collins2018}, on-chip studies of microbial organisms \cite{Zhou2017, Zhang2019}, and non-trivial electrode shapes to generate chirped, focused, and rotating acoustic waves \cite{Ding2012a, Riaud2015a, Collins2016a, Riaud2017}.

The development of effective handling of submicrometer-sized particles has been less successful. It remains a challenge to handle this in biotechnology highly important class of particles including small bacteria, exosomes, and viruses. Could these particles be handled in a controlled way, it would be of particular interest for developing new and more efficient diagnostics \cite{Liga2015}. The first steps towards acoustofluidics handling of nanometer-sized particles have been taken relying on acoustic streaming effects with both bulk acoustic waves (BAW) \cite{Antfolk2014} and SAW \cite{Mao2017}, or using seed particles to enhance acoustic trapping in BAW devices \cite{Hammarstrom2012}. However, these methods have a low selectivity. However, recently SAW devices have been developed to focusing nanoparticles \cite{Collins2017} and separation of nanoparticles  \cite{Sehgal2017, Wu2017}. In particular Sehgal and Kirby \cite{Sehgal2017} demonstrated separation between 100- and 300-nm-diameter particles on the proof-of-concept stage. To fully utilize the potential of this and similar devices, further development is necessary to increase the efficiency and sorting flow rates. Here, numerical simulations may play a crucial role, both in improving the understanding of the underlying physical acoustofluidic processes, and to ease the cumbersome development cycle consisting of an iterative series of creating, fabricating, and testing device designs.

An increasing amount of numerical studies include piezoelectric dynamics in two-dimensional (2D) models \cite{Johansson2012, Garofalo2017, Darinskii2016, Tan2010a}, but mostly the piezoelectric transducers are introduced in numeric models in the form of analytic approximations \cite{Koster2007, Zhang2018, Lei2013, Ley2017, Nama2015, Vanneste2011}, and designs are often based on a priori knowledge of the piezoelectric effect in the unloaded substrates typically applied in telecommunication. In acoustofluidic devices, the acoustic impedance of the contacting fluid is much closer to that of the substrate causing waves to behave much differently from those in telecommunications devices. It is thus prudent to include the piezoelectric effect and the coupling between the fluid and substrate in numeric models to accurately describe the device behavior. Additionally, three-dimensional (3D) simulations in the literature are scarce, but they are essential for making full-device acoustophoresis predictions as many actual acoustofluidic devices do exhibit non-trivial features in 3D due to asymmetric and intricate shapes of electrodes and channels.

In this paper, we present 3D numerical simulations taking into account the electromechanical fields of the piezoelectric SAW device, the acoustic displacement field in the attached elastic material, in which the liquid-filled microchannel is embedded, the acoustic fields inside the microchannel, as well as the resulting acoustic radiation force and streaming-induced drag force acting on microparticles suspended in the microchannel. The model is validated experimentally with devices based on the SAW device described by Sehgal and Kirby~\cite{Sehgal2017}. In \secref{model} we describe the physical model system representing the SAW device and state the governing equations, and in \secref{NumImpl} we treat the implementation of the model system in a weak-form, finite-element model. The results of the model in reduced 2D and in full 3D are presented in \secsref{Results2D}{Results3D}, and finally in \secsref{discussion}{conclusion} we discuss our findings and summarize our conclusions.

\section{The model SAW system and the governing equations} \seclab{model}

The model SAW system is shown in \figref{Models}a. Essentially, it consists of a piezoelectric lithium niobate substrate with a specific interdigitated transducer (IDT) metal-electrode configuration on the surface. On top of the substrate a microfluidic channel is defined in an elastic material, either the acoustically soft rubber polydimethylsiloxane polymer (PDMS) or the acoustically hard borosilicate glass (Pyrex).

We follow Sehgal and Kirby~\cite{Sehgal2017} and place the IDT electrodes directly underneath the microchannel and choose the periodicity of the electrode pattern to result in a SAW wavelength $\lamSAW = 80~\SImum$ and a (unloaded) resonance frequency $\fSAW = c_\mr{SAW}/\lamSAW = (3995~\SIm/\SIs)/(80~\SImum) = 49.9$~MHz. The driving electrodes are flanked by Bragg-reflector electrodes to (partially) reflect the outgoing SAWs traveling along the surface from the driving electrodes. As described in more detail in \appref{Bond}, the lattice coordinate system $X,Y,Z$ of the \lnb\ wafer is rotated the usual $38^\circ = 128^\circ - 90^\circ $  about the $x$-axis to obtain an optimal SAW configuration.

To facilitate separation of nanoparticles, the axis of the microchannel is tilted $10^\circ$ angle relative to the IDT electrodes. At both ends, the microchannel branches out in a number of side channels with vertical openings for inlet and outlet tubing. In the numerical model, this inlet/outlet structure is represented by ideally absorbing boundary conditions.

The SAW device is actuated by a time-harmonic voltage difference at frequency $f$ applied to the IDT electrodes. The corresponding angular frequency is  $\omega = 2 \pi f$.

The following formulation of the governing equations, is a further development of our previous work presented in Refs.~\cite{Skov2016, Ley2017, Skov2019} to take into account SAW in 3D models of lithium-niobate-driven ultrasound acoustics in liquid-filled microchannels.

\begin{figure}[t]
	\centering \includegraphics[clip,width=0.8\columnwidth]{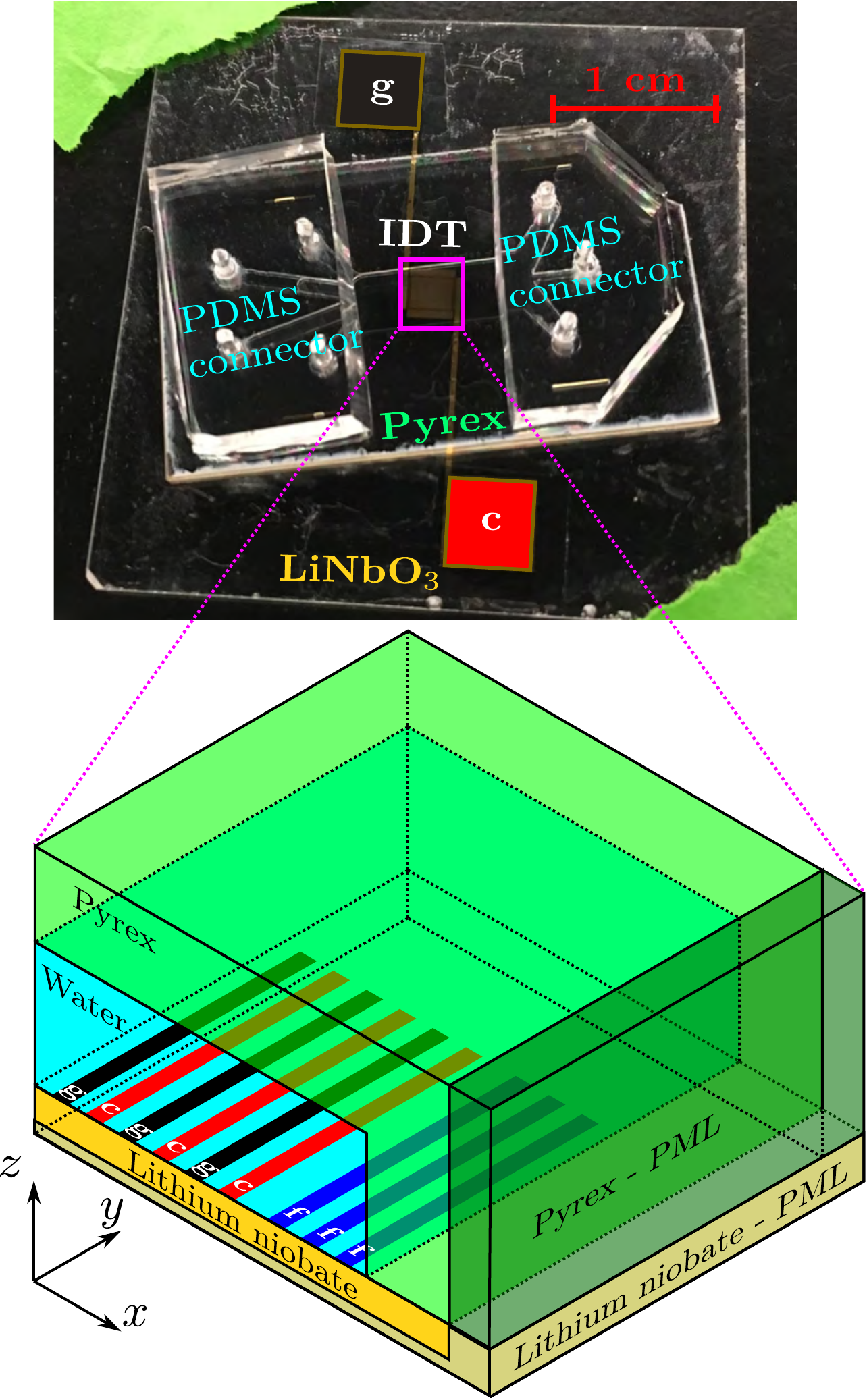}
	\caption{\figlab{Models} Experimental and numeric testing devices. (a) A testing device, similar to that of Ref.~\cite{Sehgal2017}. A wide lithium niobate base with a 24-pair interdigitated surface metal electrode (IDT) and contact pads (grounded g or ge, charged c or ce) supporting a borosilicate glass (Pyrex) slab containing an etched microchannel above the IDT. (b) 3D sketch of the numerical model containing only a 3-pair electrode (grounded g or ge: black, charged c or ce: red), and three floating electrodes (f or fe, blue).}
\end{figure}

\subsection{The Voigt notation for elastic solids}

In linear elastodynamics with the elasticity tensor $C_{iklm}$, the stress $\sigik$ and strain $\eikm$ tensors with $i,k = 1,2,3$ (or $x$, $y$, $z$) are defined in index notation as
 \bsubal
 \eqlab{epsikm} 	\eikm &= \frac12 \big(\pp_i u_k + \pp_k u_i\big), \\
 \eqlab{sigik} \sigma_{ik} &= C_{iklm} \eps_{lm} 
 \esubal
In the Voigt notation (subscript V) \cite{Auld1990}, the symmetric stress and strain double-index tensor components $\sigik=\sigma_{ki}$ and $\eikm = \epsilon_{ki}$ are organized in single-index vectors $\sigma_\alpha$ and $\epsilon_\alpha$ with $\alpha = 1,2,\ldots,6$, as,
 \bsubal
 \epsV &= \begin{pmatrix}
 \epseps_{1}\\\epseps_{2}\\\epseps_{3}\\\epseps_{4}\\\epseps_{5}\\\epseps_{6}
 \end{pmatrix} \hspace{-1mm} = \begin{pmatrix}
 \epseps_{11}\\ \epseps_{22}\\ \epseps_{33}\\2 \epseps_{23}\\2\epseps_{13}\\2\epseps_{12}
 \end{pmatrix},\quad 
 \sigV = \begin{pmatrix}
 \sigma_{1}\\\sigma_{2}\\\sigma_{3}\\\sigma_{4}\\\sigma_{5}\\\sigma_{6}
 \end{pmatrix}  \hspace{-1mm} =  \begin{pmatrix}
 \sigma_{11}\\ \sigma_{22}\\ \sigma_{33}\\\sigma_{23}\\\sigma_{13}\\\sigma_{12}
 \end{pmatrix}, \eqlab{sigIV}
 \esubal
and the stress-strain relation is written,
 \beq{VoigtStressStrain}
 \sigma_\alpha = C_{\alpha\beta} \eps_\beta,
 \eeq
where $C_{\alpha \beta}$ is the 6$\times$6 Voigt elasticity matrix. We also introduce the 3$\times$6 Voigt matrix gradient operator $\nablaV$,
 \beq{nablaV} 
 \nablaV =  \begin{pmatrix}
	\pp_x 	& 0 			& 0 			& 0 			& \pp_z 	& \pp_y \\
	0 			& \pp_y 	& 0 			& \pp_z 	& 0 			& \pp_x \\
	0 			& 0 			& \pp_z 	& \pp_y 	& \pp_x 	& 0 \\ \end{pmatrix}.
 \eeq

The equations governing the device are divided into three sets. One set is the first-order time-harmonic equations for the acoustic fields, the second set contains the steady time-averaged second-order fields, and the third set are the time-dependent equations describing the acoustophoretic motion of suspended particles.

\subsection{The time-harmonic first-order fields} \seclab{Gov1}

By construction, all first-order fields are proportional to the time-harmonic electric potential actuating the SAW device at angular frequency $\omega$. Consequently, all first-order fields are time-harmonic acoustic fields of the form  $\hat{g}(\rrr,t) = g(\rrr)\:\ee^{-\ii \omega t}$, where $g(\rrr)$ is the complex-valued field amplitude. The corresponding physical field is the real part $\re\!\big[\hat{g}(\rrr,t)\big]$. All terms thus have the same explicit time dependence $\ee^{-\ii \omega t}$, so this factor is divided out, leaving us with the governing equations for the amplitude $g$, where we for brevity suppress the spatial argument $\rrr$.

In a linear piezoelectric material with a mass density $\rhosl$ and no free charges, the solid displacement field $\uuu$ and the electric potential field $\phi$ are governed by the Cauchy equation and Gauss's law,
 \bsub
 \eqlab{CauchyGauss}
 \bal
 \eqlab{CauchyV}	\divV \sigV   &= -\rhosl \omega^2\: \uuu, \\
 \eqlab{Gauss} \div \DDD  &= 0.
 \eal
This equation system is closed by the constitutive equations relating the stress $\sigV$ and the electrical displacement $\DDD$ to the strain $\epsV$ and the electric field $\EEE$, through the elasticity matrix $\CCC$, the relative dielectric tensor $\er$, and the piezoelectric coupling matrix $\eee$,
 \eqlab{sPDP}
 \bal
 \eqlab{sP} \sigV &= \CCC \epsV - \eee\transp \EEE, \: \text{ with } \EEE = -\grad \phi, \\
 \eqlab{DP}\DDD &=  \varepsilon_{0}\er \EEE + \eee \epsV.
 \eal
 \esub
Here, $\ve_0$ is the vacuum permittivity, $\er$ is the relative permittivity tensor of the material, and superscript ``T" denotes the transpose of a matrix, see \tabref{Materials}.

For anisotropic lithium niobate, \eqsref{CauchyV}{Gauss} are turned into equations for $\uuu$ and $\phi$ by using the explicit form of \eqsref{sP}{DP} written as the coupling-matrix,
\bsub
 \beq{stress_Voigt}
 \left(\!\! \begin{array}{c}
	\sigma_{1} \\  \sigma_{2} \\  \sigma_{3} \\ \hline
	\sigma_{4} \\  \sigma_{5} \\  \sigma_{6} \\ \hline
	D_x \\ D_y \\ D_z \\
\end{array} \!\! \right)
\!\!=\!\!
\left(\!\! \begin{array}{c@{\:}c@{\:}c@{}|c@{}c@{}c@{}|c@{}c@{}c}
	C_{11} 	& C_{12} 	& C_{13} 	& C_{14} 	& 0 		& 0 		& 0				 & \text{-}e_{21}	& \text{-}e_{31} \\
	C_{12} 	& C_{22} 	& C_{23} 	& C_{24} 	& 0 		& 0 		& 0 			 & \text{-}e_{22} 	& \text{-}e_{32} \\
	C_{13} 	& C_{23} 	& C_{33} 	& C_{34} 	& 0 		& 0 		& 0 			 & \text{-}e_{23} 	& \text{-}e_{33} \\ \hline
	C_{14} 	& C_{24} 	& C_{34} 	& \!C_{44} 	& 0 		& 0 		& 0 			 & \text{-}e_{24} 	& \text{-}e_{34} \\
	0 		& 0 		& 0 		& 0 		& C_{55} 	& C_{56} 	& \text{-}e_{15}& 0 				 & 0 \\
	0 		& 0 		& 0 		& 0 		& C_{56} 	& C_{66}  	& \text{-}e_{16}& 0 				 & 0 \\ \hline
	0 		& 0 		& 0 		& 0 		& e_{15} 	& e_{16} 	& \ve_{11} 		 & 0 				 & 0 \\
	e_{21} 	& \!e_{22} 	& \!e_{23} 	& \!e_{24} 	& 0 		& 0 		& 0 			 & \ve_{22} 			 & \ve_{23} \\
	e_{31} 	& e_{32} 	& e_{33} 	& \!e_{34} 	& 0 		& 0  		& 0 			 & \ve_{23} 			 & \ve_{33}\\
\end{array} \!\! \right) \!\!\!
\left(\!\!\!\!   \begin{array}{c}
	\eps_{1} \\ \eps_{2} \\ \eps_{3}
	\\[-1.5mm] \rule{8mm}{0.15mm}\\[-0.5mm]
	\eps_{4} \\ \eps_{5}  \\ \eps_{6}
	\\[-3mm] \rule{8mm}{0.15mm}\\[-0.5mm]
	E_x  \\ E_y  \\ E_z  \\
\end{array} \!\!\!\!  \right)\! ,
\eeq

\begin{table}[t]
	\centering
	\caption{\tablab{Materials} Elasticity constants $C_{\alpha \beta}$, mass density $\rhosl$, piezoelectric coupling constants $e_{i\alpha}$ and relative dielectric constants $\varepsilon_{ik}$ of materials used in this work. \lnb\ values are defined in the global system $x,y,z$, for derivations see \appref{Bond}. Note that $C_{12} = C_{11}-2C_{44}$ for isotropic materials (Pyrex and PDMS).}
	\begin{ruledtabular}
		\begin{tabular}{crrr}
			Parameter 							& Value 		& Parameter 		&   Value     			 \\ \hline
			\multicolumn{4}{l}{\textit{\lnb}  \cite{Weis1985}}				 \rule{0mm}{1.1em}  		 \\
			$C_{11}$  	& 202.89 GPa	& $C_{12}$ 		& 72.33	GPa					\\
			$C_{13}$	& 60.17	GPa	& $C_{14}$ 		& 10.74	GPa					\\
			$C_{22}$  	& 194.23 GPa	& $C_{23}$ 		& 90.59	GPa					\\
			$C_{24}$	& 8.97 GPa	& $C_{33}$ 		& 220.29 GPa					\\
			$C_{34}$	& 8.14 GPa	& $C_{44}$ 		& 74.89	GPa					\\
			$C_{55}$	& 72.79	GPa	& $C_{56}$ 		& $-$8.51	GPa					\\
			$\rhosl$  	& $4628 \, \SIkgpcm$ & $C_{66}$	& 59.51	GPa   						 \\	
			$e_{15}$  	& 1.56 $\SIC \, \SIm^{-2}$		& $e_{16}$ 		& -4.23 $\SIC \, \SIm^{-2}$						 \\
			$e_{21}$  	& $-$1.73	$\SIC \, \SIm^{-2}$	& $e_{22}$ 		& 4.48	$\SIC \, \SIm^{-2}$					 \\
			$e_{23}$  	& $-$1.67 $\SIC \, \SIm^{-2}$	& $e_{24}$ 		& 0.14	$\SIC \, \SIm^{-2}$					 \\
			$e_{31}$  	& 1.64 $\SIC \, \SIm^{-2}$		& $e_{32}$ 		& $-$2.69	 $\SIC \, \SIm^{-2}$					 \\
			$e_{33}$  	& 2.44 $\SIC \, \SIm^{-2}$	& $e_{34}$ 		& 0.55	$\SIC \, \SIm^{-2}$					 \\	
			$\ve_{11}$  & 44.30		& $\ve_{22}$ 	& 38.08						\\
			$\ve_{23}$  & $-$7.96 	& $\ve_{33}$ 	& 34.12						\\[2mm]		
			\multicolumn{4}{l}{\textit{Pyrex} \cite{Narottam1986}} 							 \\
			$C_{11}$  	& 69.73	GPa	& $C_{12}$  	& 17.45	GPa					\\
		    $\rhosl$  	& $2230 \, \SIkgpcm$&$C_{44}$  	& 26.14	GPa  \\
			$\varepsilon$	& 4.6 & $\GammaS$ & 0.0002   							 \\[2mm]
			\multicolumn{4}{l}{\textit{PDMS} \cite{Madsen1983,Zell2007,MIT_PDMS}} 							 \\
			$C_{11}$  	& 1.13	GPa	& $C_{12}$  	& 1.11	GPa					\\
			$\rhosl$ &  $1070 \, \SIkgpcm$ & $C_{44}$  	& 0.011	GPa	 \\
			$\varepsilon$	& 2.5 & $\GammaS$ &0.0213				
		\end{tabular}
	\end{ruledtabular}
\end{table}

For isotropic elastic solids with no charges and no piezoelectric coupling $\eee = \zerovec$,
only \eqref{CauchyV} is relevant, and it becomes an equation for $\uuu$, as \eqref{sP} reduces to
 \beq{stress_Iso}
\left(\!\! \begin{array}{c}
	\sigma_{1} \\  \sigma_{2} \\  \sigma_{3} \\ \hline
	\sigma_{4} \\  \sigma_{5} \\  \sigma_{6}
\end{array} \!\! \right)
\!\!=\!\!
\left(\!\! \begin{array}{c@{\:}c@{\:}c@{}|c@{}c@{}c}
	C_{11} 	& C_{12} 	& C_{12} 	& 0 		& 0 		& 0 					\\
	C_{12} 	& C_{11} 	& C_{12} 	& 0 		& 0 		& 0 		 			 \\
	C_{12} 	& C_{12} 	& C_{11} 	& 0 		& 0 		& 0 		 			 \\ \hline
	0 		& 0 		& 0 		& C_{44} 	& 0 		& 0 		 			 \\
	0 		& 0 		& 0 		& 0 		& C_{44} 	& 0 	 \\
	0 		& 0 		& 0 		& 0 		& 0 		& C_{44}   	 \\
\end{array} \!\! \right) \!\!\!
\left(\!\!\!\!   \begin{array}{c}
	\eps_{1} \\ \eps_{2} \\ \eps_{3}
	\\[-2mm] \rule{8mm}{0.15mm}\\[-0.5mm]
	\eps_{4} \\ \eps_{5}  \\ \eps_{6}
\end{array} \!\!\!\!  \right)\!,
\eeq
\esub
with only two independent elastic constants, $C_{11}$ and $C_{44}$, because $C_{12} = C_{11}-2C_{44}$ for isotropic material.

In a fluid with speed of sound $\cfl$, mass density $\rhofl$, dynamic viscosity $\etafl$, viscous boundary layer thickness $\delta = \sqrt{\frac{2 \etafl}{\rhofl \omega}}$, viscosity ratio $\beta=\frac{\etafl^\mr{b}}{\etafl}+\frac13$, and effective damping coefficient $\GammaF = \frac{1+\beta}{2}(k_0 \delta)^2$, the first-order pressure field $\ppI$ is governed by the Helmholtz equation, and the acoustic velocity field $\vvvI$ is given by the pressure gradient,
 \bsubal
 \eqlab{Helmholtz}
 \div \big(\grad p_1\big) &= -k_c^2 p_1, \: \text{ with }
 k_c = \frac{\omega}{\cfl}\bigg(1+\ii\frac{\GammaF}{2}\bigg),\\
 \eqlab{v1def}
 \vvvI &= \frac{-\ii}{\omega\rhofl}\big(1-\ii\GammaF\big)\nablabf\pI,
 \esubal
where $k_c$ is the weakly damped compressional wave-number\cite{Bach2018}. See \tabref{Fluid} for parameter values.

\begin{table}[t]
	\centering
	\caption{\tablab{Fluid} Material parameters of water from Ref.~\cite{Muller2014}.}
	\begin{ruledtabular}
		\begin{tabular}{lcc}
			Parameter 			& Symbol & Value 							 \\ \hline
			Speed of sound 		& $\cfl$  			& 1497~$\SIm \, \SIs^{-1}$  		 \rule{0mm}{1.1em} \\
			Mass density 		& $\rhofl$  		& 997~$\SIkg \, \SIm^{-3}$  	 \\
			Dynamic viscosity	& $\etafl$  		& 0.89~m$\SIPas$		  		 	 \\
			Bulk viscosity		& $\etafl^\mr{b}$  	& 2.485~m$\SIPas$		  		 	 \\
			Compressibility     & $\kappafl$		& 452~$\SIpTPa$
		\end{tabular}
	\end{ruledtabular}
\end{table}

Turning to the boundary conditions, we introduce $\nnn$ as the normal vector for a given surface. The SAW device in \figref{Models} is actuated by a time-harmonic potential of amplitude $V_0$ on the surfaces of the charged electrodes (ce) and 0~V on the grounded electrodes (ge), respectively,
 \bsub
 \eqlab{phiBC}
 \beq{ElecPair} 
 \phi_\mr{ce} = V_0 \: \eiot \quad , \quad \phi_{\mr{ge}} = 0,
 \eeq
A given floating electrode (fe) is modeled as an ideal equipotential domain with a vanishing tangential electrical field on its surface,
 \beq{ElecSlope} 
 (\oneten - \nnn\nnn) \cdot \grad \phi_\mr{fe} = \zerovec, 
 \eeq
where $\oneten$ is the unit tensor, and $(\oneten - \nnn\nnn)$ is the usual tangent projection tensor. Note that this condition is automatically enforced on any surface with a spatially invariant Dirichlet condition applied along it. Note also that the value of the potential on each floating electrode is a priori unknown and must be determined self-consistently from the governing equations and boundary conditions.
\esub 

At a given fluid-solid interface we impose the usual continuity conditions \cite{Ley2017} with the recently developed boundary-layer corrections included \cite{Bach2018}:  the solid stress $\sigsl$ is given by the acoustic pressure $\pI$ with the addition of the boundary-layer stress, and the fluid velocity $\vvvI$ is given by the solid-wall velocity $\vvvsl = -\ii\omega\uuu$ with the addition of the boundary-layer velocity $\vvvsl - \vvv_1$,
 \bsub
 \eqlab{p1u1BC}
 \bal
 \eqlab{WeakV1} \sigsl \cdot \nnn &= - \pI\: \nnn + \ii\ks\etafl(\vvvsl - \vvv_1\big), \\
 \eqlab{Weakp1} \nnn \cdot \vvvI &= \nnn\cdot\vvv_\mr{sl} +
 \frac{\ii}{k_s}\nablabf_\parallel\cdot \big(\vvv_\mr{sl}-\vvvI\big),\\
 &\quad \text{with shear wavenumber } k_s = \frac{1+\ii}{\delta}.
 \eal
 \esub
The terms containing the shear wavenumber $k_s$ represent the corrections arising from taking the 400-nm wide, viscous boundary layer into account analytically \cite{Bach2018}.

All exterior solid surfaces facing the air have a stress-free boundary condition prescribed,
 \beq{SolidAir}
 \sigmabf \cdot \nnn  = \zerovec.
 \eeq
This is a good approximation because the surrounding air has an acoustic impedance 3 to 4 orders of magnitude lower than that of the solids causing 99.99 \% of incident acoustic waves from the solid to be reflected. Moreover the shear stress from the air is negligible.

\subsection{The time-averaged second-order fields} \seclab{Gov2}
The slow timescale or steady fields in the fluid are the time-averaged second-order velocity $\vvvII$ and pressure $\ppII$ field. These are governed by the time-averaged momentum and mass-conservation equations,
 \bsub 
 \eqlab{GovEqu2}
 \bal
 \eqlab{NS2}
 \div \sig_2 - \rho_0 \div \avr{\vvvI\vvvI} & = \zerovec, \\
 \eqlab{cont2}
 \div \big(\rho_0 \vvvII + \avr{\rho_1 \vvvI}\big) & = 0,
 \eal
where $\sig_2$ is the second-order stress tensor of the fluid
 \beq{sigmaII}
 \sig_2 = - \ppII \oneten + \eta \, \big[\grad \vvvII + \left(\grad \vvvII\right)\transp\big] + (\beta - 1) \, \eta \, (\div \vvvII) \, \oneten.\\
 \eeq
 \esub
Along a fluid-solid interface with tangential vectors $\eee_\xi$ and $\eee_\eta$ and the normal vector $\eee_\zeta = \nnn$, we use for $\vvvII$ the effective boundary condition derived in Ref.~\cite{Bach2018}. Here, the viscous boundary layer is taken into account analytically by introducing the boundary-layer velocity field  $\vvvds = \vvvsl-\vvvI$ in the fluid along the fluid-solid interface,
 \bsub
 \eqlab{BCv2}
 \bal
 \eqlab{vl2_bc_final}
 \vvvII &= \big(\AAA\cdot\een_\xi\big)\:\een_\xi + \big(\AAA\cdot\een_\eta\big)\:\een_\eta
 + \big(\BBB\cdot\een_\zeta\big)\:\een_\zeta,
 \\
 \AAA &= -\frac{1}{2\omega}  \Re \bigg\{ \vvvdsC_1\scap\grad\Big(\frac{1}{2}\vvvds_1-\ii\vvvsl\Big)
 -\ii\vvvsl^* \scap\grad \vvvI
 \\ \nn
 &\hspace*{15mm} +\bigg[\frac{2-\ii}{2}\grad\scap\vvvdsC_1
 +\ii\Big(\grad\scap\vvvsl^*-\pp_\zeta v_{1\zeta}^*\Big)\bigg]\vvvds_1 \bigg\},
 \\
 \BBB &=
 \frac{1}{2\omega} \Re\Big\{ \ii\vvv_1^*\scap\grad\vvvI\Big\},
 \eal
 \esub
where the asterisk denotes complex conjugation.

\subsection{Acoustophoresis of suspended particles} \seclab{Gov3}
To predict the acoustophoretic motion of a dilute suspension of spherical micro- and submicrometer-sized particles in the fluid of density $\rhofl$, compressibility $\kappafl$, and viscosity $\etafl$, we implement a particle tracing routine in the model. We consider Newton's second law for a single spherical particle of radius $a_\mr{pt}$ and density $\rho_\mr{pt}$ moving with velocity $\vvvP$ under the influence of gravity $\vec{g}$, the acoustic radiation force $\FFFrad$ \cite{Settnes2012}, and the Stokes drag force $\FFFdrag$ \cite{Bruus2011} induced by acoustic streaming of the fluid,
 \bsub
 \eqlab{ParticleMotion}
 \bal
 \eqlab{Newton2}
 \frac{4\pi}{3} a_\mr{pt}^3  & \rho_\mr{pt}\frac{\dd\vvvP}{\dd t} =
 \rho_\mr{pt} \vec{g} + \FFFrad + \FFFdrag,\\
 \eqlab{Frad}
 \FFFrad &= -\frac{4}{3} \pi a^3 \Big[\kappafl \avr{(f_0 p_1)\grad p_1} -\frac32 \rhofl \avr{(f_1 \vvvI)\cdot \grad \vvvI} \Big],\\
 \eqlab{Fdrag}
 \FFFdrag &= 6 \pi a \etafl \big(\vvvII - \vvvP\big).
 \eal
Here, $f_0 = 0.444$ and $f_1 = 0.034$ are the monopole and dipole scattering coefficients of the suspended particles at 50 MHz, where the values are for polystyrene micro- and nanoparticles in water \cite{Karlsen2015}. When studying different particle sizes it is convenient to introduce the radiation force density $\fffrad$ as
 \beq{def_fffrad}
 \fffrad = \frac{3}{4\pi a^3}\:\FFFrad.
 \eeq
 \esub

By direct time integration of \eqref{Newton2} applied  to a set of particles initially placed on a square grid, the acoustophoretic motion of the particles can be predicted and compared to the experimentally observed one. We note that gravity effects are negligible as $\rho_\mr{pt} \vec{g}~\ll~\kappafl \avr{(f_0 p_1)\grad p_1}$.

\begin{table}[b]
		\caption{\tablab{dim} Dimensions of the numeric 2D and 3D models.}
\begin{ruledtabular}
	\begin{tabular}{lcccc}
		Parameter           &      Symbol     &         2D          &         3D          &   Unit   \\ \hline
		Device depth  ($y$) &  $L_\mr{sl}$    &          -          &        1200         & $\SImum$ \\
		Solid height  ($z$) &  $H_\mr{sl}$    &       40-1000       &         500         & $\SImum$ \\
		Solid width   ($x$) &  $W_\mr{sl}$    &         200         &        80           & $\SImum$ \\
		Channel height      &  $H_\mr{fl}$    &       50-200        &         50          & $\SImum$ \\
		Channel width       &  $W_\mr{fl}$    &        3500         &         900         & $\SImum$ \\
		Piezo height        &  $H_\mr{pz}$    &       100-500       &         300         & $\SImum$ \\
		PML length          &  $L_\mr{PML}$   &         80          &         80          & $\SImum$ \\
		Electrode depth  ($y$) & $L_\mr{el}$  &          -          &        400          & $\SImum$ \\
		Electrode height ($z$) & $H_\mr{el}$  &         0.4         &         0.4         & $\SImum$ \\
		Electrode width  ($x$) & $W_\mr{el}$  &         20          &         20          & $\SImum$ \\
		Electrode gap       &   $G_\mr{el}$   &         20          &         20          & $\SImum$ \\
		SAW wavelength      &   $\lamSAW$     &         80          &         80          & $\SImum$ \\
		No.\ of electrode pairs & $n_\mr{el}$ &         24          &          4          &    -     \\
		No.\ of reflectors  &  $n_\mr{rf}$    &         0-6         &          0          &    -     \\
		Actuation frequency &      $f_{0}$    &        30-60        &         50          &   MHz    \\
		Driving voltage     &      $V_{0}$    &        1            &         1           &    V     \\
		Degrees of freedom  & $n_\mr{DOF}$    & $\mathcal{O}(10^5)$ & $\mathcal{O}(10^6)$ &    -     \\
		Memory requirements &        $R$      &  $\mathcal{O}(10)$  & $\mathcal{O}(10^3)$ &    GB
	\end{tabular}
\end{ruledtabular}
\end{table}

\section{Numerical implementation} \seclab{NumImpl}

Inspired by our previous experimental work, Sehgal and Kirby \cite{Sehgal2017}, we study the SAW test system shown in \figref{Models} with actuating electrodes and Bragg-reflector electrodes placed directly underneath the microchannel. The parameter values used in the numerical simulation are listed in \tabref{dim}, and a sketch of the vertical cross section of the test system is shown in \figref{Electrodes}. Note that the SAW wavelength $\lamSAW$ is set by the IDT electrode geometry as $\lamSAW = 2(W_\mr{el}+G_\mr{el})$. We study microcavities defined in either acoustically soft PDMS, see \figref{Electrodes}(a), or the acoustically hard borosilicate glass (Pyrex), see \figref{Electrodes}(b), and we perform numerical simulation in both 2D and 3D.

\begin{figure}[t]
	\includegraphics[clip,width=0.95\columnwidth]{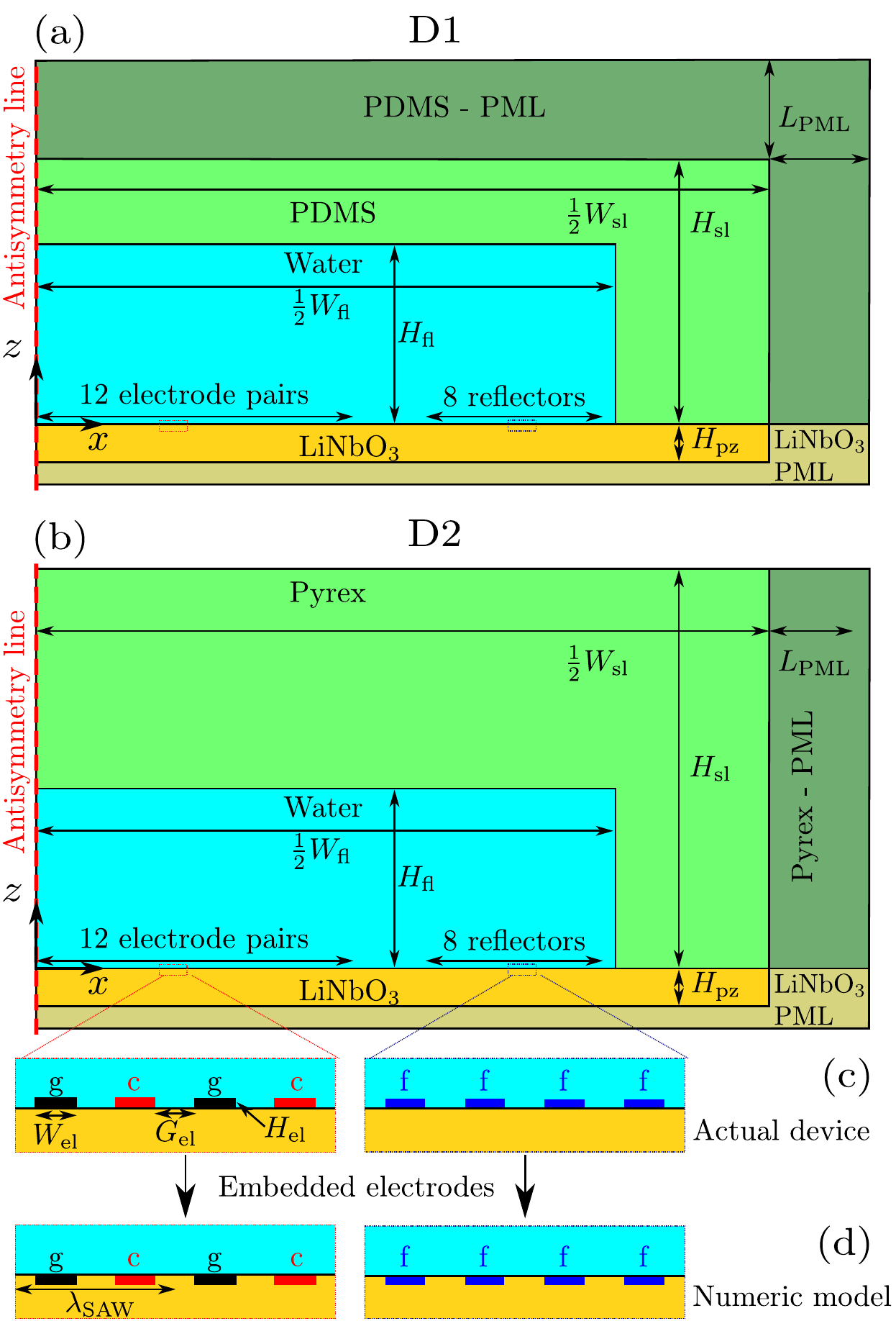}
	\caption{\figlab{Electrodes} The vertical 2D cross section of the numeric model and illustration of the embedded electrodes used in simulations, with (a) a highly attenuating, low-reflection polymer PDMS lid as used in Ref.\ \cite{Sehgal2017}, and (b) a stiff, acoustically reflecting Pyrex glass lid. (c) The twelve pairs of grounded (g, black) and charged (c, red) electrodes, as well as the floating (f, blue) electrodes, are all included in their entire height, but (d) lowered into the lithium niobate (yellow) to level with the substrate. Note that $\lamSAW = 2(W_\mr{el}+G_\mr{el})$.}
\end{figure}

Following the procedure of our previous numerical simulations~\cite{Ley2017, Skov2019}, the coupled governing equations from Sections \ref{sec:Gov1}-\ref{sec:Gov3} are implemented in the finite-element-method software COMSOL Multiphysics~5.3a \cite{COMSOL53a}, using the weak-form partial differential equation interface ``\textit{PDE Weak Form}" in the mathematics module. For a given driving voltage $V_0$, actuation frequency $f$, and angular frequency $\omega = 2\pi f$ specified in the actuation boundary condition~\eqrefnoEq{ElecPair}, the numeric model is solved in three sequential steps: (1) the first-order equations~\eqsrefnoEq{CauchyGauss}{Helmholtz} presented in \secref{Gov1} for the pressure $p_1$, displacement $\uuu$, and electric potential $\phi$, together with the corresponding boundary conditions~\eqrefnoEq{phiBC}-\eqrefnoEq{SolidAir}; (2) the steady second-order streaming velocity $\vvv_2$ in \secref{Gov2} governed by \eqsrefnoEq{GovEqu2}{BCv2}, where time-averaged products of the first-order fields appear as source terms; and (3) the acoustophoretic motion of suspended test particles in \secref{Gov3} found by time integration of \eqref{ParticleMotion}. As in previous works \cite{Ley2017}, we have performed convergence analyses of the model to verify that the model converges towards a single solution as the mesh size decreases.

Simulations of the full 3D model are time and computer-memory consuming. Therefore, part of the analysis has been performed on 2D models to study the resonance behavior of the device and the acoustic radiation force in the vertical $y$-$z$ plane normal to the electrodes in the horizontal $x$-$y$ plane. In these simulations, presented in \secref{Results2D}, it is possible to model a cross-section of the device to scale. To investigate effects that have non-trivial behavior in full 3D, such as the acoustic streaming and the acoustophoretic motion of suspended particles presented in \secref{Results2D}, we must perform full 3D modeling. However, in this case, the extended computer memory requirements has necessitated a  scale down of the model. The parameters for the 2D and 3D simulations are listed in \tabref{dim}.

\subsection{Perfectly matched layers}
\seclab{PML}
We reduce the numeric footprint of the model by implementing perfectly matched layers (PMLs) in the model as described by Ley and Bruus \cite{Ley2017}: Large passive domains surrounding the acoustically active region are replaced by much smaller domains, in which PMLs act as ideal absorbers of out-going acoustic waves thus completely removing reflections. In contrast to Ref.~\cite{Ley2017}, the PMLs in the present model are functions of all three spatial coordinates.

In the small surrounding domains, the PMLs are implemented in the weak-form governing equations by a complex-valued coordinate transformation of the spatial derivatives $\pp x_i$ and integral measures $\dd x_i$ appearing,
 \bsubal
 \pp x_i \rightarrow \pp \tilde{x}_i &= \frac{1}{1 + \ii \,s(\rrr)}\:\pp x_i \eqlab{PMLdd}, \\
 \dd x_i \rightarrow \dd \tilde{x}_i &= \big[1 + \ii \,s(\rrr)\big] \: \dd x_i \eqlab{PMLid},\\
 s(\rrr) &= k_\mr{PML}\sum_{i=x,y,z} \frac{(x_i-x_{0i})^2}{L^2_{\mr{PML},i}}\: \Theta(x_i-x_{0i}),
 \esubal
where $s(\rrr)$ is a real-valued function of position. Here, $s(\rrr)$ is given for the specific case shown in \figref{Electrodes} with a PML of width $L_{\mr{PML},i}$ in the three coordinate directions  $i = x,y,z$ placed outside the region $x<x_0$, $y<y_0$, and  $z<z_0$, $\Theta(x)$ is the Heaviside step function ($= 1$ for $x>0$, and 0 otherwise), and $k_\mr{PML}$ is an adjustable parameter for the strength of the PML absorbtion. The bottom PML in the niobate substrate is used because SAWs decay exponentially in the depth on the scale of the wavelength, whereas the top and side PMLs are used to mimic attenuating in respective materials over large distances.

\subsection{Symmetry planes}\seclab{Symmetry}
As in previous numeric works \cite{Muller2015,Ley2017}, we use an antisymmetry line to reduce the numerical cost of our 2D models. The antisymmetry line is realized by boundary conditions on the solid displacement, the electric potential, and the fluid pressure along the line,
\bsubal
\pp_x u_{x} &= 0 \\
u_{z} &= 0 \\
\phi &= \frac12 V_0 \\
p_1 &= 0
\esubal

We check these conditions against the values along the device centerline in a 2D simulation for a fully symmetric device and observe that they are in good agreement.

In 3D we cannot use symmetry planes, as the device is manifestly asymmetric due to the $10^\circ$~angle between the IDT and the walls of the microchannel.

\subsection{Embedded electrodes}\seclab{Thin film}
In the actual device, the $400$-nm thick electrodes protrude into the fluid domain. In our numeric model we simplify the device by submerging them into the substrate to form a planar solid-fluid interface as shown in \figref{Electrodes}. Thereby the fluid-solid interface has no sharp corners, at which singularities appear in the numeric gradients. Furthermore, the planar interface mitigates the need for an enormous number of mesh elements ranging from nm to $\SImum$ in the fluid domain, which would either lower the element quality greatly or add massive computational costs. This reduction in model complexity is justified by the height of the electrodes being less than 1~\% of the channel height and having no influence on the pressure acoustics of the system. On the other hand, we cannot completely neglect the electrodes, because jumps in acoustic impedance between the metal electrodes and the niobate substrate cause partial reflections of SAWs running along the substrate. Thus we choose to keep but submerge the electrodes.

\section{Experimental methods}
\seclab{Experiment}

To validate the numerical models, we have performed experiments on two type of devices listed in table IV, namely microchannels defined in slabs of either PDMS (D1) or Pyrex (D2) bonded on top of the lithium niobate substrate equipped with the IDT and Bragg reflectors. The PDMS device (D1) is fabricated by standard photolithography techniques listed in our previous work~\cite{Sehgal2017}. The Pyrex device (D2) is fabricated by glass microfabrication techniques, briefly described in the following. A microchannel of desired dimensions is wet-etched in a borosilicate glass wafer by 49\% hydrofluoric (HF) acid using a multilayered mask of chrome, gold, and SPR220 photoresist. The input and output ports of the microchannel are obtained from the laser cutting of glass. The bonding between glass microchannel and lithium niobate substrate is achieved by coating a 5~$\SImum$ layer of SU-8 epoxy on the surface of lithium niobate. The microchannel is gently placed on the uncured SU-8 and the epoxy is baked following standard steps. The SU-8 outside the microchannel region is selectively crosslinked to achieve bonding and the SU-8 inside the microchannel region is dissolved away with a developer, thus obtaining a Pyrex lid microchannel on top of the lithium niobate substrate (D2).
The devices are tested with 1.7-$\SImum$-diameter fluorescent polystyrene particles (Polysciences, Inc.) that are suspended in de-ionized water (18.2 M$\Omega$/cm, Labconco WaterPro PS) containing 0.7\% (w/v) Pluronic F-127 to prevent particle aggregation. The particle solution is injected into the microchannel after priming the devices with 70\% ethanol solution to avoid the formation of air bubbles. An ultrasound field is set up in the devices by applying an RF signal at desired frequency to the IDT with a HP 8643A signal generator and an ENI 350L RF power amplifier. The acoustophoretic motion of the tracer particles are visualized on a fixed-stage, upright fluorescent microscope (Olympus BX51WI) with a digital CCD camera (Retiga 1300, Q Imaging). The images are acquired with Q-Capture Pro 7 software and post processed in ImageJ. The electrical impedance of the devices is measured directly from an impedance analyzer (Agilent 4395A).

\begin{table}[t]
\caption{\tablab{DevicesD1D2} The devices D1 and D2, used in the experimental validation of the numerical model, differs by the choice of lid. The other parameters of D1 and D2 are listed in \tabref{dim}.}
\begin{ruledtabular}
	\begin{tabular}{lcc}
		Device           &      Lid material      &  Lid thickness  \\ \hline
		D1               & PDMS &   15~mm  \rule{0em}{3ex}  \\
		D2, see \figref{Models}(a) & Pyrex &  0.45~mm
	\end{tabular}
\end{ruledtabular}
\end{table}

\section{Results of the 2D modeling}
\seclab{Results2D}

In the following, we compare the results of the 2D modeling in the vertical $x$-$z$ with experiments carried out on the two devices D1 and D2 listed in \tabref{DevicesD1D2}. Such a comparison is reasonable because the low channel height of 50~$\SImum$ implies an approximate translation invariance along the $y$-axis spanning the length (aperture) 2400~$\SImum$ of the IDT electrodes, as seen in the 3D geometry of \figref{Models}. Also the variation along the $x$ axis given by the width $20~\SImum$ of the individual electrodes, and the periodicity $\lamSAW = 80~\SImum$ the IDT, are much smaller than IDT aperture along $y$ axis. We can therefore obtain a reasonable estimate of the electrical and acoustical response of the device, by just considering the 2D domain in the vertical  $x$-$z$ plane shown in \figref{Electrodes}.

\begin{figure}[b]
 \includegraphics[width=1\columnwidth]{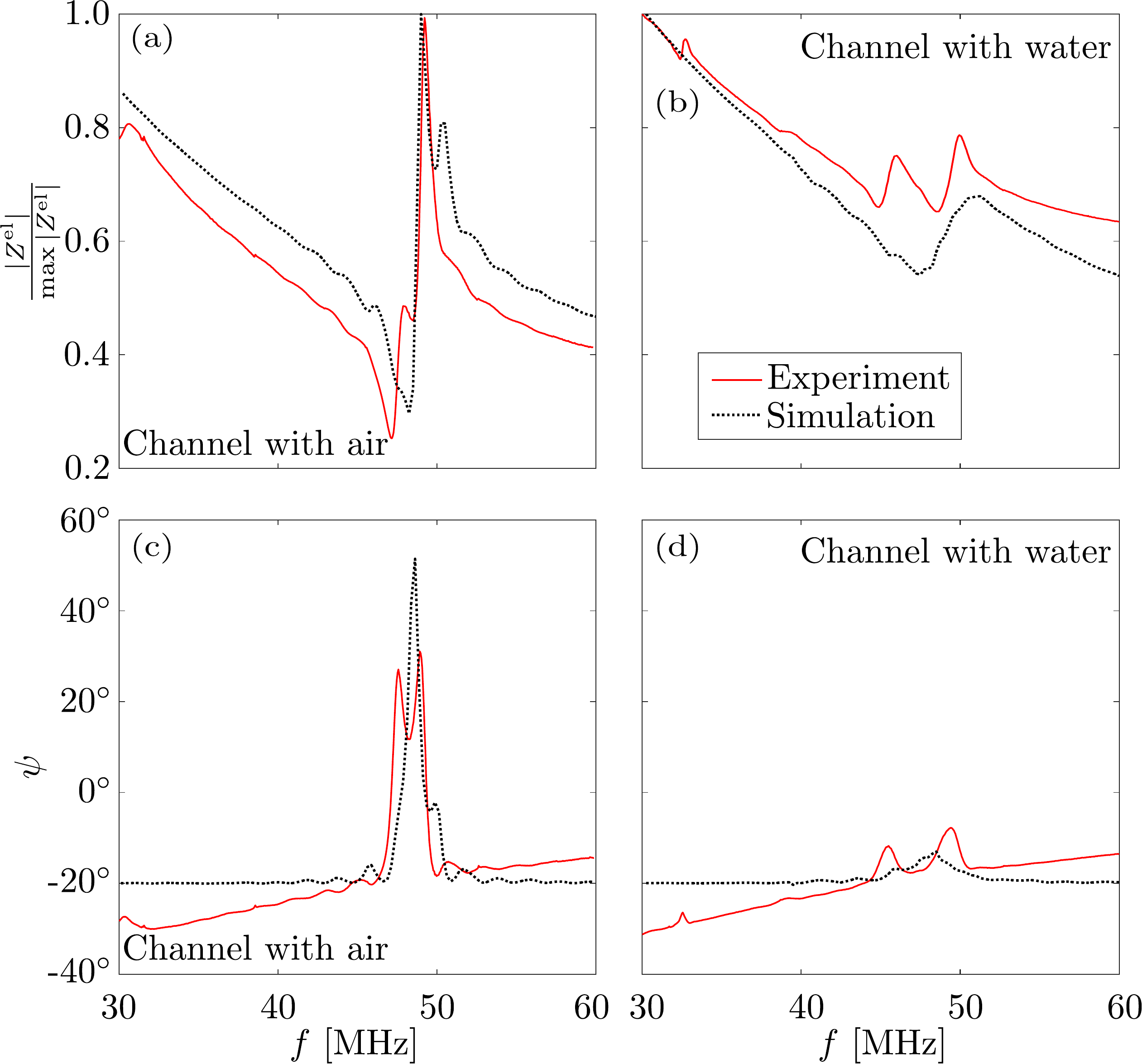}
 \caption{\figlab{RC} Line plots of the normalized magnitude $\big|\Zel\big|$ and phase $\psi$ of the electrical impedance $\Zel$, as functions of frequency determined by experiment (full red line) and by numerical simulation (dotted black line). The measurements and simulations are carried out for a microchannel containing either vacuum or deionized water.}
\end{figure}

\subsection{Electrical response} \seclab{ElecResponse}
As a first validation of the model, we study the electrical impedance
 \beq{ZelDef}
 \Zel = \big|\Zel\big|\:e^{-\ii \psi}= \frac{V_0}{I},
 \eeq
in terms of the driving voltage $V_0$ and the complex-valued current $I$ through the device, because this quantity is relatively easy to obtain both in simulation and in experiment. We compare model predictions of the magnitude $\big|\Zel\big|$ and phase $\psi$ of the impedance with the experimentally measured counterparts.

In the model, we compute $\Zel$ from the time-harmonic dielectric polarization density $\PPP$ and the corresponding polarization current $\JJJ_\mr{pol}$ in the lithium niobate substrate, which we treat as an ideal dielectric without free charges,
 \bsub
 \bal
 \eqlab{Polarization}
 \PPP &= \DDD - \varepsilon_0 \EEE,\\
 \eqlab{Currentdensity}
 \JJJ_\mr{pol} &= -\ii \omega \PPP.
 \eal
The total current $I$ through the device is given by the surface integral of $\JJJ_\mr{pol}$,
over one of the charged electrodes with potential $\phi_\mr{ce} = V_0$ and surface $\pp\Omega_\mr{ce}$,
 \beq{Current}
 I = \int_{\pp\Omega_\mr{ce}} \JJJ_\mr{pol} \cdot \nnn \, \dA.
 \eeq
The modulus $\big|\Zel\big|$ and phase angle $\psi=\arg(\Zel)$ are thus
 \beq{ZelFinal}
 \big|\Zel\big| = \left|\frac{V_0}{I}\right|, \qquad
 \psi = \arg\left(\frac{V_0}{I}\right).
 \eeq
 \esub

\begin{table}
	\caption{\tablab{RC} Measured and simulated values of the frequencies $f$ near the ideal (unloaded) frequency  $\fSAW = 49.9$~MHz, where $|\Zel(f)|$ and $\psi(f)$ have local minima and maxima in Pyrex device D2.}
	\begin{ruledtabular}
		\begin{tabular}{cccc}
			Extremum  &  $f_\mr{exp}$  &  $f_\mr{num}$  &    Relative error\\
					& [GHz]& [GHz] & [\%] \\ \hline
			A & $47.35$  & $48.25$ &   1.9  \rule{0em}{3ex} \\
			B & $48.05$  & $49.00$ &   2.0   \\
			C & $49.40$  & $50.25$ &   1.7   \\
			D & $48.65$ &  $47.50$ &   1.7   \\
			E & $50.00$ &  $50.75$ &   1.7   \\
			F & $47.70$  & $48.75$ &   2.2   \\
			G & $49.10$  & $50.00$ &   1.8   \\
			H & $45.50$  & $46.00$ &   1.1   \\
			I & $49.40$  & $48.50$  &   1.8
		\end{tabular}
	\end{ruledtabular}
\end{table}

In \figref{RC}(a) and (b), we compare the values of  $|\Zel|$ computed by \eqref{ZelFinal} for our 2D model with those measured on Pyrex device D2 of \figref{Models}(a) and \tabref{DevicesD1D2} for microchannels with air or with DI water. The numerical simulation predicts correctly the value of the resonance observed near 48~MHz in the experiments. As shown in \tabref{RC}, the relative difference between computed and measured values of the frequencies $f$, where $|\Zel(f)|$ and $\psi(f)$ have local minima or maxima, is about 2~\% or less. We also see that simulation also predicts the monotonically decreasing background signal for $|\Zel(f)|$ before and after the resonance relatively well for both an air- and water-filled microchannel. However, the simulation fails to predict the correct ratio of the resonance peak heights.

For the phase $\psi$ shown in \figref{RC}(c) and (d), the simulation predicts the resonance frequencies correctly, but fails to predict the monotonically increasing background signal. By adding external stray impedances to our 2D model to simulate the surrounding 3D system, it is however possible to generate a slant in the phase curves by fitting the values of these stray impedances. We do not show these results as they are descriptive and not predictive in nature.

\subsection{Wall material: hard pyrex versus soft PDMS} \seclab{PDMSPyrex}
Our previous device~\cite{Sehgal2017} features a soft PDMS polymer lid, as is commonly used due to the ease of fabrication and handling. However, the acoustic properties of PDMS are far from ideal: its impedance is nearly equal to that of water (20~\% lower) and the attenuation is about two orders of magnitude larger than that of the boundary layer in water. In the following, we therefore simulate the acoustic properties of the device D1 with a PDMS lid and contrast them with those of device D2 with a much stiffer Pyrex lid, using the two models shown in \figref{Electrodes} and \tabref{DevicesD1D2}. Compared to water, the acoustic impedance of Pyrex is 8.3 times larger and its attenuation 10 times smaller.

We study by numerical simulation the acoustic fields of device D1 and D2 near the ideal (unloaded) frequency  $\fSAW = 49.9$~MHz. By locating the maximum of the average acoustic energy in the water-filled channel plotted versus the actuation frequency $f$ (not shown), we determine the (loaded) resonance frequency $f_\mr{res}$ of the two devices to be $f_\mr{res}^\mr{D1} = 47.75$~MHz and $f_\mr{res}^\mr{D2} = 46.50$~MHz, respectively. In \figref{p1uPDMSPY} we show line plots along the height ($z$ direction) and across the width ($x$ direction) of numerically simulated acoustic fields for these two devices.

In  \figref{p1uPDMSPY}(a) and (b) is shown the magnitude $|u_z|$ of the $z$ component of the acoustic displacement $\uuu$, which in water is defined through acoustic velocity \eqref{v1def} as $\vvv_1 = -\ii\omega\: \uuu$, along a vertical cut-line through the entire device. In D1, $|u_z|$ has the characteristics of a traveling wave emitted from the SAW substrate (maximum amplitude), traversing the water with little reflection (a small oscillation amplitude), and being absorbed in the PDMS lid (decaying amplitude). In contrast, $|u_z|$ in D2 has the characteristics of a standing  wave localized in the water channel with reflections from the surrounding solids: huge oscillations in the water domain with minima close to zero and an amplitude exceeding that in the emitting substrate and the receiving lid. We also notice that in the stiff Pyrex the attenuation is weak, and that the wave is reminiscent of a standing wave between the water interface below the lid and the air interface above. The corresponding  acoustic energy flux density $\SSS_\mr{ac} = \avr{p_1 \vvv_1}$ in both systems is non-zero and predominantly vertical, but with a much larger amplitude in D1 compared to D2.

In  \figref{p1uPDMSPY}(c) is shown the the magnitude $|\uuu|$ of the acoustic displacement $\uuu$ along horizontal cut-lines following the top ($z=H_\mr{fl}$) and the bottom ($z=0$) of the water channel across the region containing the IDT. In both devices the periodicity of the IDT electrodes is clearly seen, but the amplitude in the nearly-standing wave case of D2 is 2-3 times larger than in the traveling wave case of D1. Moreover, it is seen that the acoustic waves dies out faster in D1 than in D2 away from the IDT region. The tiny oscillations in the PDMS lid (green curve) for $x < -12 \lamSAW$ stems from the minute transverse wavelength $\sim 11~\SImum = 0.13~\lamSAW$ in PDMS.

\begin{figure}[t]
	\includegraphics[width=1\columnwidth]{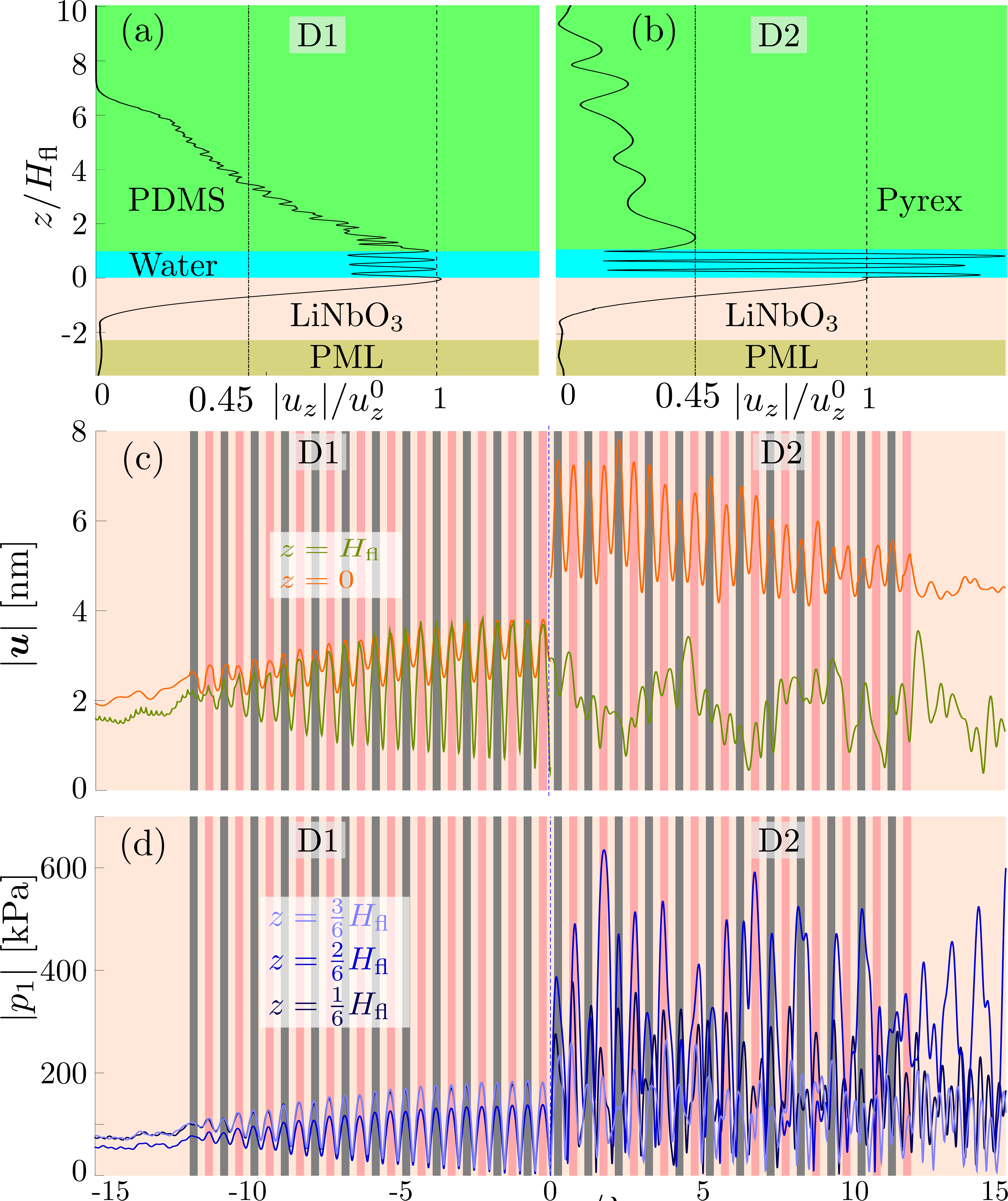}
	\caption{\figlab{p1uPDMSPY} The amplitude of the displacement  $|\uuu|$ and the pressure amplitude $|p_1|$ in the PDMS-lid device D1 and in the Pyrex-lid device D2 at their respective resonance frequencies $f_\mr{res}^\mr{D1} = 47.75$~MHz  and  $f_\mr{res}^\mr{D2} = 46.50$~MHz at $V_0 = 1$~V. (a) Line plot of the $z$ component $|u_z|$ along the vertical line $x= W_\mr{el}$ (the center of the middle electrode) from the bottom of the substrate (beige), through the water (blue), to the top of the PDMS lid (green). (b) As in (a), but for the Pyrex-lid device D2. (c)  Line plot of $|u|$ along the top ($z=H_\mr{fl}$) and the bottom ($z=0$) of the channel in D1 ($x<0$) and D2 ($x>0$). The dark gray and pink rectangles for $-12 < \frac{x}{\lamSAW} < 12$ represent the IDT electrodes. (d) As in panel (c), but for $|p_1|$ along the horizontal lines at $z/H_\mr{fl} = \frac36, \frac26, \frac16$ inside the channel.}
\end{figure}

In  \figref{p1uPDMSPY}(d) is shown the the magnitude $|p_1|$ of the acoustic pressure $p_1$ in the water along the horizontal cut-lines $z/H_\mr{fl} = \frac16, \frac26, \frac36$. Here the traveling versus standing wave nature of the two devices mentioned above, is prominent: In D1, $|p_1|$ is nearly independent of the height, and its envelope amplitude is steadily decaying from 90 to 55~kPa from the center to the edge of the IDT region. In contrast, $|p_1|$ has large amplitude fluctuations as a function of the horizontal position $x$ and for the three vertical $z$ positions. Moreover, $|p_1|$ does not decay away from the the IDT. Clear, $p_1$ in the water channel of D2 is dominated by reflections between the solid-water interfaces. This observation can be quantified by the the standing wave ratio, $\mr{SWR} = {\max(|\ppI|)}/{\min(|\ppI|)}$ that describes the ratio of standing to traveling waves in a given field. In an ideal resonator and an ideally transmitting system, $\mr{SWR} = \infty$ and 1, respectively. Here, we find $\mr{SWR(D2)} = 12.7$ and $\mr{SWR(D1)} =1.3$. These numbers underlines the good acoustic properties of the water-Pyrex systems compared to the bad one of the PDMS system. The ratio of the SWR numbers is 9.8, almost equal to the impedance ratio 10.5, which emphasizes the nearly perfect vertical energy flux density $\SSS_\mr{ac}$ discussed above, as the impedance extracted from the properties of a plane wave with a vertical incident on a planar surface.

\subsection{Acoustophoresis} \seclab{2DARF}

Whereas we have not made experimental validation of the above simulation results for the acoustic fields $p_1$ and $\uuu$, we compare in the following the experimentally observed acoustophoretic motion at the SAW resonance frequency $\fSAW$ of microparticle suspensions in the water-filled microchannel, with that obtained by numerical simulation in our 2D model. The central experimental and numerical results are shown in \figref{2Dfradv2}, in the left column for the PDMS-lid device D1 and in the right column for the Pyrex-lid device D2. In the Supplemental Material~\footnote{The Supplemental Material at [URL] contains four animations of the numerically simulated acoustophoresis of 0.1- and 1.7-$\SImum$-diameter particles in device D1 and D2, respectively, corresponding to \figref{2Dfradv2}(c) and (g).} are shown four animations of the acoustophoresis in \figref{2Dfradv2}(c) and (g) of 0.1- and 1.7-$\SImum$-diameter particles in device D1 and D2.

\begin{figure*}[t]
	\includegraphics[width=2\columnwidth]{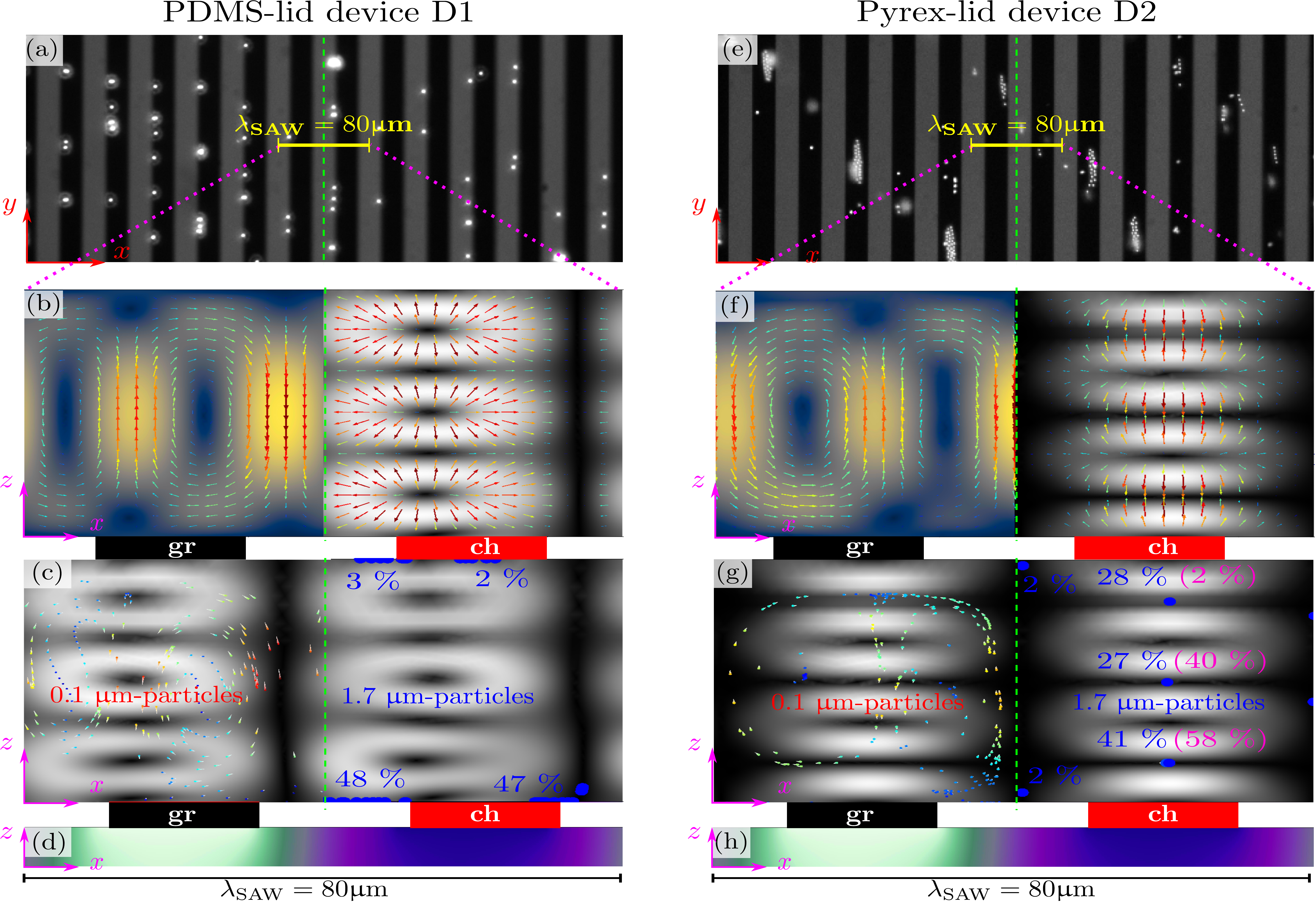}
	\caption{\figlab{2Dfradv2} Microparticle acoustophoresis in experiments and in simulations for actuation frequency $\fSAW = 49.9~\SIMHz$ and driving voltage $V_0 = 4.35~\SIV$, rescaling the simulation from 1 to 4.35~V.
(a) Top-view photograph ($x$-$y$ plane) of the center region of the IDT array in device D1, where suspended 1.7-$\SImum$-diameter fluorescent polystyrene particles (white) are focused above the edge of each metal electrode (black).
(b) Numerical simulations in the vertical $x$-$z$ plane over a single electrode pair ($6\lamSAW < x < 7\lamSAW$, the yellow line in panel (a)) in the fluid domain of device D1 with (to the left) a color plot of the magnitude $|\vvv_2|$ [from 0 (blue) to $66~\SImum/\SIs$ (yellow)] of the streaming velocity $\vvv_2$, and (to the right) a gray-scale plot of $|\fffrad|$ [from $0$ (black) to $0.4~\SIpN/\SIum^3$ (white)] of the acoustic radiation force density $\fffrad$. Superimposed are colored vector plots of $\vvv_2$ [from 0 (blue) to $66~\SImum/\SIs$  (red)] and of $\fffrad$ [from 0 (blue) to 0.4~$\SIpN/\SIum^3$ (red)].
(c) Color-comet-tail plot of the simulated acoustophoretic motion of 247 0.1-$\SImum$-diameter spherical polystyrene particles (to the left), superimposed on the gray-scale plot of $|\fffrad|$ from panel (b), 0.5~s after being released from initial positions in a regular 13$\times$19 grid to the left of the green-dashed centerline. Similarly for 1.7-$\SImum$-diameter particles to the right. The comet tail indicates the direction of the velocity with length and color from $0$ (dark blue) to $66~\SImum/\SIs$  (orange) representing the speed. The percentages indicate the portion of particles accumulating in these final positions: the blue set for a homogeneous initial particle distribution, and the purple set for an inhomogeneous initial particle distribution created by 3~min of sedimentation.
(d) Color plot in the vertical $x$-$z$ plane below a single electrode pair $5\lamSAW < x < 6\lamSAW$ of the numerically simulated electric potential $V$ from $-4.35$ (light cyan) to $4.35~\SIV$ (purple) in the lithium niobate substrate. The width and $x$-position of the grounded and charged electrodes in the IDT-pair are represented by the black (ge) and red (ce) rectangles, respectively.
(e-h) Same as in (a-d) but for Pyrex-lid device D2, and in (f) the gray-scale for  $|\vvv_2|$ is from 0 (blue) to $76~\SImum/\SIs$  (yellow) and $|\fffrad|$ from $0$ (black) to $7.4~\SIpN/\SIum^3$~(white).}
\end{figure*}

In \figref{2Dfradv2}(a) and (e) we observe that the suspended 1.7-$\SIum$-diameter particles in D1 focus on the edges of the electrodes, whereas in D2 they mainly focus along the center line of each electrode. This difference in acoustophoretic focusing is caused solely by choice of lid material and its thickness. Already in \figref{p1uPDMSPY}, we saw how the change from the PDMS lid to the Pyrex lid led to a change from a predominantly traveling wave, to a nearly standing wave in the $z$ direction. As a consequence, both the pressure and its gradients in device D1 are smaller than those in D2, and from \eqref{Frad} follows that the acoustic radiation force  $\FFFrad$ changes significantly.

This change in $\FFFrad$ per particle volume,  named $\fffrad$ in \eqref{def_fffrad}, is shown as the vector and gray-scale plots for device D1 and D2 in the right half of \figref{2Dfradv2}(b) and (f), respectively. Compared to D2 having  $|\fffrad| = 7.4~\SIpN/\SIum^3$, the magnitude $|\fffrad| = 0.4~\SIpN/\SIum^3$ is 18 times smaller in D1, and $|\fffrad|$ is more smeared out (even smaller gradients). Both force fields have a three-period structure along the vertical $z$ axis, reflecting that $H_\mr{fl} \approx \frac32 \cfl/\fSAW$. In D1, the center of the force-field structure is displayed relative to the center of the electrode, whereas in D2 it is above the electrode center. Moreover, whereas $\fffrad$ has four less-marked, unstable nodal planes in D1 at $z/H_\mr{fl} = 0, \frac13, \frac23, 1$, it has three well-defined, stable ones in D2 at $z/H_\mr{fl} = \frac16, \frac36, \frac36$.

The corresponding streaming velocity field $\vvv_2$ in D1 and D2 is shown as the vector and color plots in the left half of \figref{2Dfradv2}(b) and (f). The streaming appears strikingly equal both in magnitude ($66~\SImum/\SIs$ for D1 and $76~\SImum/\SIs$ for D2), shape and topology, but again with the center of the pattern in D1 shifted slightly away from the electrode center. The reason for this resemblance in $\vvv_2$ stems from the energy flux density $\SSS_\mr{ac}$, which in both devices points (nearly) vertically up along the $z$ axis above the electrodes, and is weak in between. As the (Eckart) streaming is proportional to $\SSS_\mr{ac}$ \cite{Eckart1948}, even in microcavities \cite{Skov2019}, the streaming moves upward due to $\SSS_\mr{ac}$ above the electrodes and downward by recirculation between the electrodes. $\SSS_\mr{ac}$ has nearly the same amplitude in D1 and D2 because, although the acoustic field in D2 is much larger than in D1, it is mostly a standing wave with zero energy flux density, and the little part that is a traveling wave in D2 that carries the energy flux density, is nearly of the same magnitude as the traveling wave that constitutes the main part of the weaker acoustic field in D1.

According to Newton's second law~\eqrefnoEq{ParticleMotion}, the above-mentioned properties of the acoustic radiation force density $\fffrad$ and streaming velocity field $\vvv_2$ governs the observable acoustophoretic motion of suspended particles. In \figref{2Dfradv2}(c) and (g), as well as in the Supplemental Material~\cite{Note1}, is shown the results of simulating such motion for 0.1- and 1.7-$\SImum$-diameter polystyrene beads in both D1 and D2, 0.5~s after starting from an initial homogeneous distribution (blue points and percentage numbers). The motion of the large particles is dominated by the radiation force~\cite{Muller2012}, so the different focusing of these particles seen in the right half of \figref{2Dfradv2}(c) and (g) is explained in terms of $\fffrad$: Because $\fffrad$ has no stable nodal planes in D1, all particles accumulate the floor or the ceiling of the channel, and most of them (98~\%) are pushed to the regions above the electrode gaps as indicated by the vector plot in the right half of \figref{2Dfradv2}(c). In contrast, the stable nodal planes of $\fffrad$ in D2, \figref{2Dfradv2}(g) right half, guides 96~\% of the particles into the three stable points above the electrode center, with 41~\%, 27~\%, and 28~\% at $z/H_\mr{fl} = \frac16 = 0.17$, $\frac36 = 0.50$, and $\frac56 = 0.83$, respectively. If we instead, as in the experiments described below, allow for a sedimentation time of 3~min before turning on the acoustics, the distribution of the focused particles changes to  58~\%, 40~\%, and 2~\% at $z/H_\mr{fl} = \frac16 = 0.17$, $\frac36 = 0.50$, and $\frac56 = 0.83$, respectively.

The acoustophoretic motion of the small 0.1-$\SIum$-diameter particles are dominated by the Stokes drag from the streaming field $\vvv_2$, see the left side of \figref{2Dfradv2}(c) and (g) and the videos in the Supplemental Material~\cite{Note1}. The simulation shows that the particles do not settle in fixed positions but follow oblong paths in the vertical plane similar in shape to the large streaming rolls spanning the entire height of the channel with an upwards motion over the electrodes and downwards in between electrodes, see \figref{2Dfradv2}(b) and (f). In D1, $\FFFrad$ is so small that it plays essentially no role. In D2, however, $\FFFrad$ is stronger and superposes with $\FFFdrag$ to govern the acoustophoretic motion. This superposition of forces is similar to the analysis presented by Antfolk \etal~\cite{Antfolk2014}, but whereas in their system the nanoparticles spirals towards the point at the center of a single flow roll, the nanoparticles above a single electrode in D2 are focused into the center line of each of the two flow rolls shown in  \figref{2Dfradv2}(f). The location of these center lines are defined by the vertical and horizontal nodal lines of $\fffrad$ represented by the black regions at the electrode gaps $x/\lamSAW = \frac{n}{2}$ and at the stable nodal planes $z/H_\mr{fl} = \frac16, \frac56$, respectively, in the gray-scale plot of \figref{2Dfradv2}(f) and (g).

Most of these theoretical predictions are validated by experiments. After loading the particle suspension in to the device, it takes about 3~min for the fluid to come to rest, during which time the 1.7-$\SIum$-diameter particles sediment slowly. This partial sedimentation shifts the homogeneous particle distribution downwards, so that the particle distribution is inhomogeneous when the acoustic field is turned on. In the experiments on PDMS-lid device D1, the large 1.7-$\SIum$-diameter particles are observed to accumulate at the floor and the ceiling in the regions between the electrodes, and the small 0.1-$\SIum$-diameter particles are observed to circulate in broad streaming rolls. In contrast, in the experiments on the Pyrex-lid device D2, the large  particle are seen to accumulate above the center of the electrodes near two planes, 36~\% of them at $z=(15\pm5)~\SImum = (0.3\pm0.1)H_\mr{fl}$ and 64~\% of them at $z=(30\pm5)~\SImum =  (0.6\pm0.1)H_\mr{fl}$. Here, the uncertainty is estimated from the optical focal depth in the setup. These numbers are in fair agreement with the simulation results mentioned above and shown in \figref{2Dfradv2}(g) (purple numbers). Finally, the observed acoustophoretic focusing time of 0.1~s matches the theoretical predictions.

\section{Results of the 3D modeling}
\seclab{Results3D}
In this section we address the more realistic, but also more cumbersome simulations in 3D for the Pyrex-lid device D2. Even given our access to the High Performance Computing clusters at the DTU Computing Center (HPC-DTU) \cite{HPCDTU}, we cannot simulate the entire chip shown in \figref{Models}(a). Whereas we keep the correct dimensions in the height, we scale down the width and length to both be around 1~mm. The 3D model geometry is shown in \figref{3DOut} with the detailed parameter values listed in \tabref{dim}. In this reduced geometry, the IDT contains only 4 electrode pairs and no Bragg reflectors. Although the model is down-sized in two of the three dimensions, it still contains all the main components of a acoustofluidic SAW device: A first step, in which the piezo-electric device, the IDT electrodes, the elastic lid, and the microchannel with the fluid and its viscous boundary layer, are combined in the calculation of the electrically induced acoustic fields. A second step, in which the acoustic radiation force and the acoustic streaming velocity are computed, and used in the governing equation predict the acoustophoretic motion of suspended spherical particles.

\subsection{The acoustic fields and radiation force}
\seclab{3Dacoustics}

The 3D model shown in \figref{3DOut} contains 4.6 million degrees of freedom. The calculation was distributed across 80 nodes on the HPC-DTU cluster and took 14 hours to compute. The first result is that the computed pressure and displacement fields $p_1$ and $\uuu$ in 3D are both qualitatively and quantitatively similar to the ones computed in the 2D model. For vertical slice planes parallel to the $x$-$z$ plane and place near the center of the IDT at $y = \frac12 L_\mr{sl}$, the agreement is of course better than for those near the edge of the IDT near $y = \frac12 (L_\mr{sl}\pm L_\mr{el})$, but in all cases we find the period-3 structure of $|p_1|$ along the $z$ direction seen in \figref{p1uPDMSPY}(b). Likewise, for the acoustic radiation force density, we recover the period-3 structure in $|\fffrad|$ seen in \figref{2Dfradv2}(f) and for the particle focusing points in \figref{2Dfradv2}(g). The experimental observation of this vertical focusing is thus validating this point in our 3D model.

\begin{figure}[t]
\includegraphics[width=0.95\columnwidth]{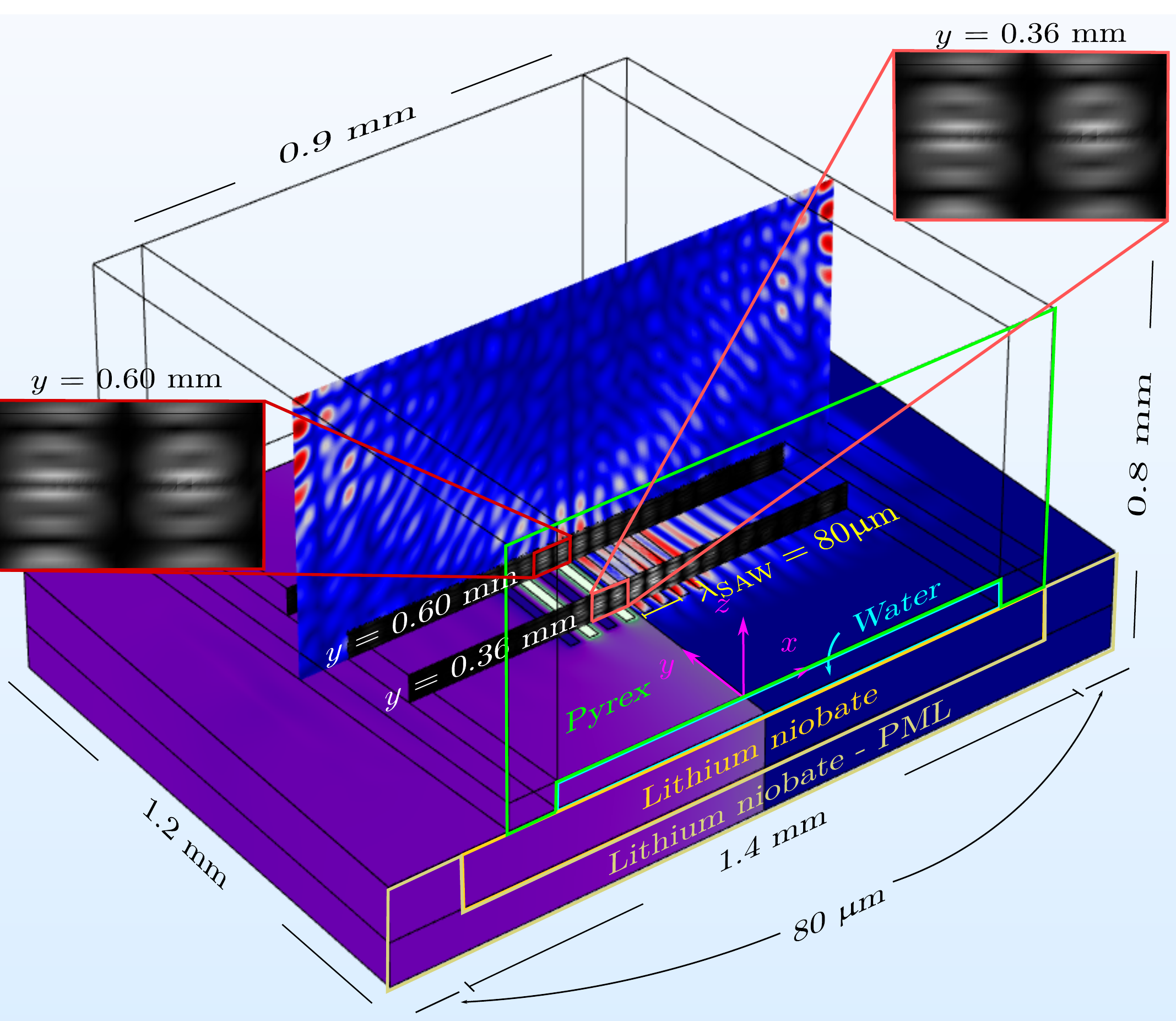}
\caption{\figlab{3DOut} A 4.4~MDOF simulation of a mm-sized Pyrex-lid device D2 in 3D actuated at $\fSAW = 50$~MHz. A surface plot of the electric potential $V$ [from $-4.35$ (purple) to $4.35~\SIV$ (light cyan), rescaled from $V_0 = 1$~V] in the piezoelectric substrate, combined with a slice plot at $y=\frac12 L_\mr{sl}$ of the acoustic pressure magnitude $|p_1|$ [from 0 (black) to 566~kPa (yellow)] in the channel and the magnitude of the displacement $|\uuu|$ [from 0 (blue) to 0.05~nm (red)] in the surrounding Pyrex.}
\end{figure}

\subsection{The acoustic streaming rolls}
\seclab{StreamRoll}
The streaming-dominated, in-plane acoustophoretic motion of 0.75-$\SImum$-diameter particles suspended in the device is used in \figref{3Dstreaming} to compare our model predictions to observed particle motion. As shown in \figref{3Dstreaming}(a), the experimentally observed particle motion in Pyrex-lid device D2 at the edges of the IDT electrodes is dominated by streaming rolls in the horizontal $x$-$y$ plane. We compare this motion with the streaming velocity field $\vvvII$ calculated using the 3D model and shown in \figref{3Dstreaming}(b). Although the model only includes a mm-sized sub-region of the experimental device, the same streaming pattern is evident in both the model device and in the experimental device. The agreement in terms of direction, position, and magnitude is good, albeit with small differences. In both the simulation and in the experiment, the centers of the streaming rolls are located at the edges of the electrodes, with clockwise-circulating flows. Similar to the 2D streaming pattern in \figref{2Dfradv2}, the observed horizontal streaming rolls are a combination of a recirculating flow and an energy flux density, here perpendicular to and away from the IDT array. The streaming velocity in D2 near the right edge of the blue rectangular region shown in \figref{3Dstreaming}(a) and (b) is measured in the 24-electrode-pair device to be $\sim200~\SImum/\SIs$ and in the simulated 4-electrode-pair device to be $\sim20~\SImum/\SIs$, or $\sim120~\SImum/\SIs$ if multiplied by the ratio of the number of electrode pairs, 24/4.

\begin{figure}[t]
		\includegraphics[width=0.96\columnwidth]{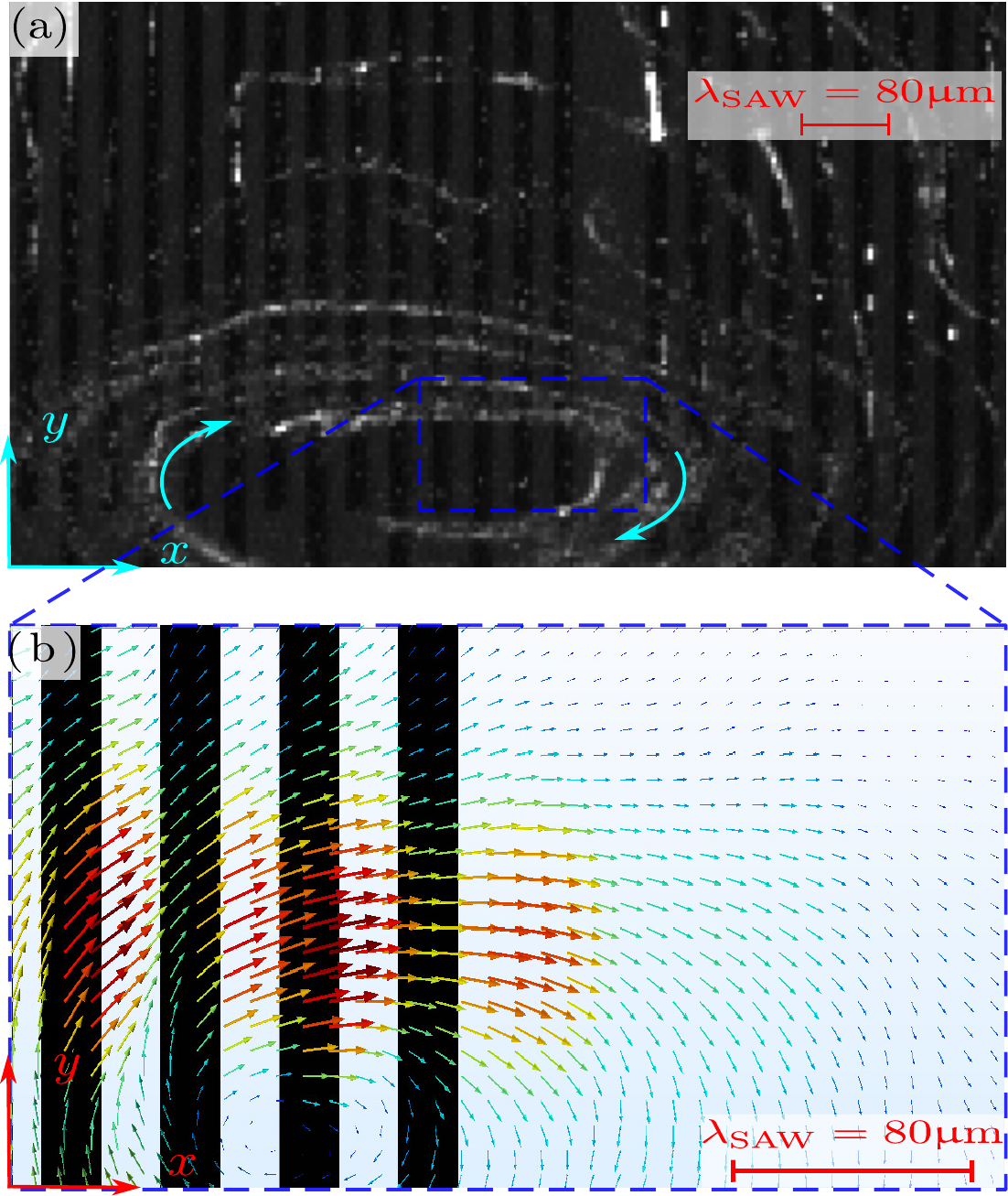}
		\caption{\figlab{3Dstreaming} Acoustic streaming in the horizontal $x$-$y$ plane of Pyrex-lid device D2. (a) Experimental top view of device D2 containing suspended 0.75-$\SImum$-diameter polystyrene particles (white), actuated at 50~MHz with $V_0 = 4.35~\SIV$. Arrows (cyan) indicate the flow direction, and the blue dashed rectangle indicates the area shown in (b). (b) Colored arrow plot of the simulated streaming velocity field $\vvvII$ [from 0 (blue) to $66~\SImum/\SIs$ (red)] in the 3D model actuated as in panel (a). The black stripes represent the electrodes.}
\end{figure}

\section{Discussion} \seclab{discussion}
By comparing our model simulations to measurable quantities, we find that the model can  predict the overall electrical and acoustophoretic behavior of the two types of SAW-devices D1 (PDMS lid) and D2 (Pyrex lid) fairly well. For the electrical response of the device we see a good agreement between the trends near resonance of the predicted and measured values of the electrical impedance, although the predicted values are obtained in an ideal 2D model neglecting stray impedances. The predicted acoustophoretic focusing of the 1.7-$\SImum$-diameter polystyrene particles at the ceiling and floor above the edges of the electrodes in D1, and at 1/6 and 3/6 of the channel height above the center of the electrodes in D2, agrees well with experimental observations.

An interesting feature of the model is the three half-wave resonance excited vertically in the Pyrex-lid device D2. It highlights the importance of careful consideration of the material selection for acoustofluidic devices, to fit with the desired purpose of the device. Because a PDMS lid is an acoustically soft material with an acoustic impedance $\Zac$ similar to that of water ($\Zac_\mr{PDMS}=1.19~\SIMRayl, \Zac_\mr{H_2O}=1.49~\SIMRayl$), most of the energy in an acoustic wave in water impinging on the water-PDMS interface is transmitted into the PDMS, where it dissipates into heat. Only a small fraction of the energy is reflected back into the fluid. As illustrated in \figref{p1uPDMSPY},  by replacing the PDMS lid of the device in Ref.~\cite{Sehgal2017} with an acoustically hard ($\Zac_\mr{Py}=12.47~\SIMRayl$) Pyrex lid, 78.6\% of the wave energy is theoretically reflected back into the fluid domain at the channel lid, compared to the 10.9\% in a PDMS lid. The resonance build-up in the microchannel is further enhanced, as the height of the channel sustains three half-waves at the resonant frequency of the IDT, $f_\mr{res}  = \frac{c_\mr{SAW}}{\lamSAW} = \frac{\cfl}{3 \lambda_\mr{Wa}} $. This resonance behavior is very similar to the integer-half-wave resonances common in BAW devices, whereas the beneficial energy localization at the surface of the SAW is still retained. Thus, the energy loss and heat generation occurring in the piezo-electric substrate in BAW devices is mitigated in this device whereas strong microchannel resonances can be achieved, when using a Pyrex lid in an IDT-inside SAW design. Considering this the terms 'BAW' and 'SAW' seem inadequate when describing acoustofluidic devices, as the actuation scheme of the piezo-electric transducer alone does not suffice to describe the resonance behavior of a device. A more descriptive feature of a device is the nature of the wave field in the fluid, because we show the main factor determining acoustophoresis in the SAW is the difference between traveling and standing wave fields in the fluid.

In acoustofluidic focusing devices, a strong streaming flow is often detrimental to the desired application, as they tend to counteract the radiation force by pulling small particles away from the nodes. In the Pyrex-lid device, however, the vertical part of the streaming enhances particle focusing, as it pulls particles from areas with weak radiation force into the lower node of the acoustic radiation force, increasing the focusing efficiency.

\section{Conclusion}
\seclab{conclusion}

We have presented a 3D model, and implemented it in the finite-element software COMSOL Multiphysics, for numerical simulation of SAW-devices taking into account the piezo-electric substrate, the IDT metal electrodes, the elastic solid defining the microchannel, the water in the microchannel as the viscous boundary layer of the water. With such simulations, we are able to decrease the gap between the systems that we can model and those used in actual experiments. This work thus brings us closer to the point, where numerical simulation can guide rational design of acoustofluidic devices.

To push acoustofluidic devices closer to medical application, the development of novel device designs beyond the proof-of-concept stage is vital. We have presented a close-to-scale numeric model of an acoustofluidic SAW device by expanding on previous model experiences \cite{Ley2017, Skov2019} and the recently developed effective-boundary-layer theory \cite{Bach2018}. With this we have captured the inner workings of a non-trivial device. The model includes the linear elasticity of the defining material, the scalar pressure field of the microchannel fluid and the piezoelectricity of the lithium niobate substrate.

Using the numeric model, we illustrate the impact that the material selection in acoustofluidic chips has on acoustophoretic performance. Based on the numerically predicted acoustic fields, we propose design improvements over the previous design \cite{Sehgal2017}, consisting primarily of substituting the original PMDS lid with a Pyrex lid. According to our model, the new lid leads to higher energy densities and more uniform particle focusing. This causes the chip to build up strong resonances in a standing wave field, similar to those in a BAW device. Furthermore, we have used our model to predict the electrical response of the a 2D model of the system, the acoustophoretic focusing of particles suspended over the IDT area of the device, and the streaming motion within devices. For each of these comparison parameters we have found an agreement between predictions and experiments.

Despite our focus on a specific device design in this manuscript, the model can handle a much wider class of acoustofluidic devices. We have a developed a model that can be reshaped to simulate any BAW or SAW device design of well-characterized piezoelectric transducers, Newtonian fluids, and isotropic and anisotropic linear solids.

In future work it would be prudent to improve the model accuracy by including the temperature field to account for the thermal dependence of material parameters, particularly the fluid bulk and dynamic viscosities. To implement the temperature field one must account for the various sources of thermal generation in terms of mechanical losses and viscous dissipation described in \cite{Hahn2015}, which requires a good knowledge of the damping properties of each component of the device.

\section{Acknowledgements}
This work is partially supported by NSF CBET-1605574, NSF CBET-1804963, and  NIH PSOC- 1U54CA210184-01. The device fabrication is performed in part at the Cornell Nanoscale Facility (CNF), which is supported by the National Science Foundation (Grant ECCS-1542081).

\appendix
\section{Bond and rotation matrices} \seclab{Bond}
The elasticity, coupling and permittivity properties of mono-crystalline lithium niobate are listed in \citeref{Weis1985} for a Cartesian material coordinate system $X,Y,Z$ defined as shown in \figref{Crystal}(a). The $Z$-axis is oriented in the growth direction, the $X$-axis is the normal to one of the three mirror planes, and the $Y$-axis follows from the right-hand rule, placing it within the mirror plane the $X$-axis is normal to. The device in this manuscript, however, is manufactured on a wafer of the more commonly used \lnb. These are wafers of lithium niobate cut from a single crystal so that the positive surface normal forms a 128$^\circ$ angle with the material $Y$-axis. In our model, we define a coordinate system $x,y,z$ with the $x$-axis coinciding with the material $X$-axis, the $z$-axis normal to the wafer surface and the $y$-axis determined by the right-hand rule. This global coordinate system coincides with the material coordinate system rotated an angle $\theta = 128 ^\circ - 90 ^\circ = 38^\circ$ counter-clockwise around the $X$-axis, as shown in \figref{Crystal}(b).

\begin{figure}[t]
\begin{center}
\includegraphics[width=0.9\columnwidth]{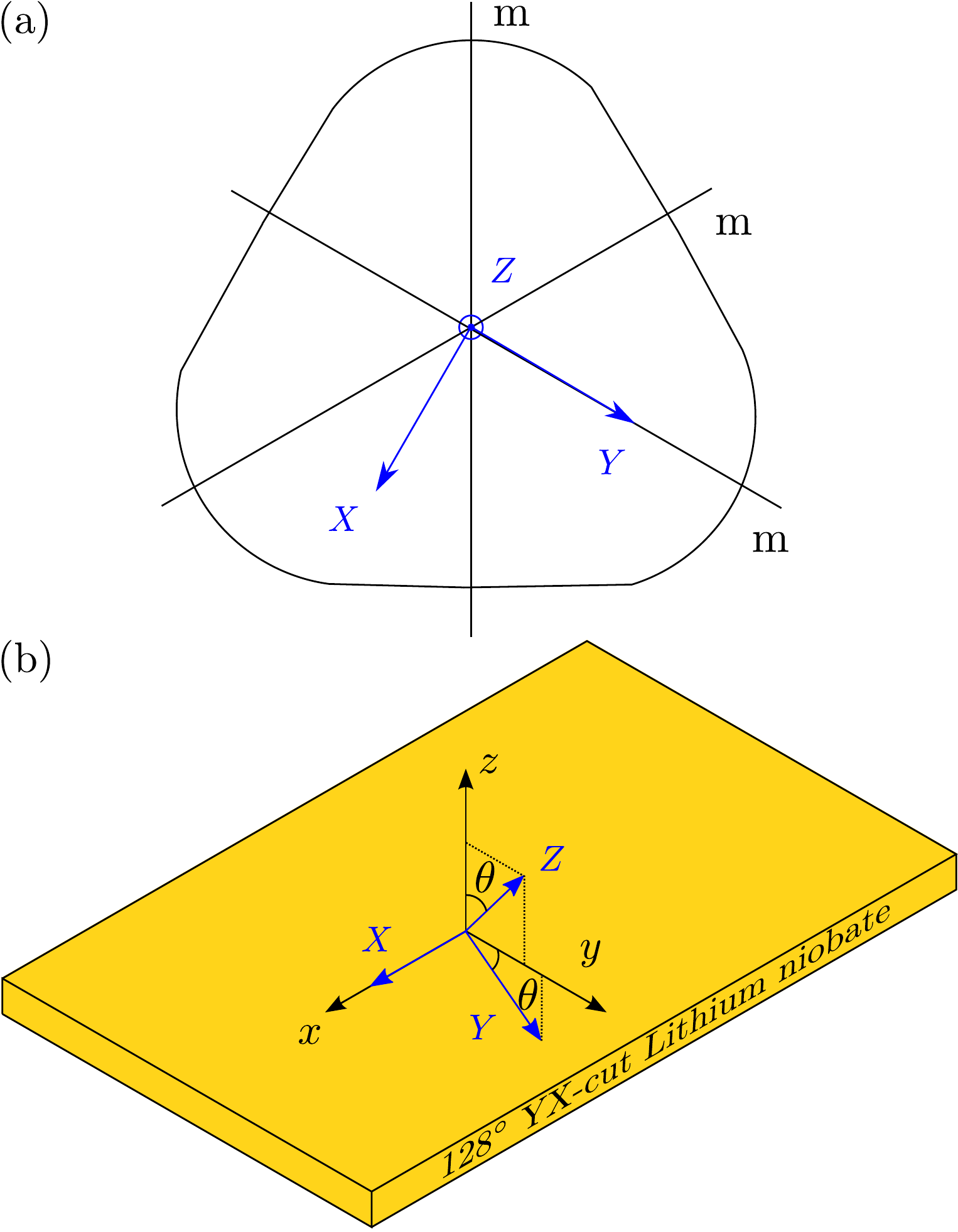}
\caption{\figlab{Crystal} (a) Top-view sketch of the material coordinate system $X,Y,Z$ in mono-crystalline, hexagonal lithium niobate with three mirror planes m. (b) \lnb\ chip showing the global coordinate system $x,y,z$ rotated counter-clockwise $\theta = 128^\circ - 90^\circ = 38^\circ$ around the $X$-axis relative to the material coordinate system $X,Y,Z$.}
\end{center}
\end{figure}

In the following, we define the matrix operations necessary to determine the material parameters in the global system $x,y,z$ from the values known in $X,Y,Z$.

In the usual Cartesian notation exists a matrix $\RRR$ transforming a 3$\times$1 vector $\PPPM$ expressed in material coordinates $X,Y,Z$ to the vector $\PPPG$ expressed in terms of a global coordinate system $x,y,z$.

\bal
\PPPG = \RRR\PPPM, \eqlab{GenTrans}
\eal

whereas 3$\times$3 matrices are transformed as
\bal
\erG = \RRR\erM\RRR\inv. \eqlab{erTrans}
\eal

For 6$\times$1 vectors in Voigt notation similar matrices called Bond matrices $\MMM_s$ transform stress vectors $\sigVM$ expressed in material coordinates into the same stress in terms of the global coordinate system $\sigVG$
\bal
\sigVG = \MMMs\sigVM, \eqlab{sigVTrans}
\eal
and similarly to \eqref{erTrans} 6$\times$6 matrices are transformed as
\bal
\cccG = \MMMs\cccM\MMMs\transp \eqlab{ElTrans}
\eal
It is important to note that Voigt notation stress and strain vectors do not transform alike and two transformation matrices exist in Voigt notation $\MMMs \neq \MMM_\epsilon$. Hence, the transformation rules deviate slightly from those in 3$\times$3 matrices.

Finally, 3$\times$6 matrices such as the coupling tensor, $\eee$ can be transformed using a rotation matrix and Bond matrix.
\bal
\eeeG = \RRR\eeeM\MMMs\transp \eqlab{eTrans}
\eal

Mathematically, a positive rotation $\theta$ degrees about the material $X$-axis is obtained by the rotation ${\RRR_x(\theta)}$ and Bond  $ \MMM_{\sigma,x}(\theta)$ matrices

\bal
{\RRR_x(\theta)} &= \begin{pmatrix} 1 & 0 & 0 \\ 0 & \ct & \st\\ 0 & -\st & \ct \end{pmatrix}, \\
 \MMM_{\sigma,x}(\theta) &=
		\begin{pmatrix}
	1 & 0 				& 0 			& 0 	& 0 	& 0 	\\
	0 & \ct^2 			& \st^2 		& 2\ct\st  & 0 	& 0 	\\
	0 & \st^2 			& \ct^2 		& -2\ct\st & 0 	& 0 	\\
	0 & -\ct\st 	& \ct\st 	& \ct^2-\st^2 	&  0 	& 0  	\\
	0 & 0  				& 0 			& 0 	& \ct 	& \st 	\\
	0 & 0 				& 0 			& 0 	& -\st 	& \ct
\end{pmatrix},
\eal
using $\ct$ and $\st$ as shorthand for $\cos(\theta)$ and $\sin(\theta)$ respectively.

%
%



%

\end{document}